\documentclass[aps,amsmath,amssymb,reprint]{revtex4-1}
\usepackage{hyperref}
\usepackage{graphicx}
\usepackage{dcolumn} % align table columns on decimal point
\usepackage{color}
\usepackage{grffile}
\usepackage{bm} % bold math

\newcommand{\phij}{\phi_{\rm J}}
\definecolor{apsblue}{rgb}{0.18,0.19,0.57}
\definecolor{darkblue}{rgb}{0.2,0.1,0.5}
\definecolor{darkgreen}{rgb}{0.1,0.6,.1}
\definecolor{darkred}{rgb}{0.7,0.0,.1}

\hypersetup{
 a4paper,
 colorlinks,
 citebordercolor={0.3 0.8 .1},
 linkbordercolor={0.9 0.2 .1},
 urlbordercolor={0.6 0.6 0.8},
 citecolor=apsblue,
 urlcolor=apsblue,
 linkcolor=apsblue,
 pdfpagemode={UseNone},
 pdfauthor={},
 pdftitle={},
}

\graphicspath{{figures/}:{./}}

\begin{document}

\title{Exploring the jamming transition 
over a wide range of critical densities}

\author{Misaki Ozawa}

\author{Ludovic Berthier}

\author{Daniele Coslovich}

\affiliation{Laboratoire Charles Coulomb, 
Universit\'e de Montpellier, CNRS, Montpellier 34095, France}

\begin{abstract}
We numerically study the jamming transition of frictionless
polydisperse spheres in three dimensions.
We use an efficient thermalisation algorithm for the 
equilibrium hard sphere fluid and generate amorphous jammed packings 
over a range of critical jamming densities that is about three times 
broader than in previous studies. 
This allows us to reexamine a wide range of structural 
properties characterizing the jamming transition.
Both isostaticity and the critical behavior 
of the pair correlation function hold over the entire range of 
jamming densities. At intermediate length scales, 
we find a weak, smooth increase of bond orientational order.
By contrast, distorted icosahedral structures grow rapidly with increasing the 
volume fraction in both fluid and jammed states.
Surprisingly, at large scale we observe that denser jammed states show 
stronger deviations from hyperuniformity, suggesting that 
the enhanced amorphous ordering inherited from the equilibrium fluid 
competes with, rather than enhances, hyperuniformity. Finally, 
finite size fluctuations of the critical jamming density 
are considerably suppressed in the denser jammed states, indicating
an important change in the topography of the potential energy landscape. 
By considerably stretching the amplitude of the critical ``J-line'', 
our work disentangles physical properties at the 
contact scale that are associated with jamming criticality, 
from those occurring at larger length scales, 
which have a different nature.
\end{abstract}

\maketitle

\section{Introduction}
\label{sec:introduction}

Granular materials such as grains, 
pinballs, and large colloids flow when some external forces are applied. 
However, when the volume fraction of these particles is increased above a certain value, particle motion is no longer allowed and these systems become 
amorphous solids.
This mechanical transition corresponds to a jamming transition
because it occurs in the absence of thermal fluctuations. 
The critical volume fraction of this critical ``J-point'' is 
$\phij$, and the associated critical properties 
have been intensively studied in the last few 
decades~\cite{liu1998nonlinear,torquato2010jammed,van2009jamming,liu2010jamming}.
In particular, structural properties of jammed states have been examined over a wide range of length scales from microscopic to macroscopic, providing a starting point for understanding the mechanical and rheological properties of jammed systems~\cite{o2003jamming,wyart2005effects,olsson2007critical,van2009jamming,lerner2012unified}.

In practice, the structural properties of jammed systems can be sorted 
out by the typical length scale or wave number $k$ that is being 
probed. First, the {\it contact scale}, corresponding to $\delta \to 0$, 
where $\delta$ is the typical gap between particles for $\phi < \phij$, or to the typical overlap between particles for $\phi>\phij$.
The corresponding wave number is $k \sim 1/\delta \to \infty$. 
A remarkable property at the contact scale is isostaticity, which 
implies that the average number of contacts per particle, $Z$, 
becomes twice the number of spatial dimensions exactly 
at $\phi=\phij$. In addition,
a critical power law behavior characterizes the pair distribution
$g(r)$ near 
contact at $\phij$~\cite{o2003jamming,van2009jamming,donev2005pair,charbonneau2012universal}.
The isostatic nature of the system is at the core of theoretical descriptions 
of the jamming transition~\cite{van2009jamming,liu2010jamming,goodrich2016scaling}.
Second, one can consider distances at the {\it neighbor scale}, corresponding
to $k \sim 2\pi/\overline{\sigma}$, where $\overline{\sigma}$ is the averaged particle diameter.
The major goal at this scale is to properly characterize the local 
geometry of amorphous configurations~\cite{clusel2009granocentric,ogarko2014communication,rieser2016divergence}, to possibly define 
appropriate order parameters for the jamming 
transition~\cite{torquato2000random,morse2014geometric}.
The role of local crystalline and icosahedral order in monodisperse packings has been discussed before~\cite{Anikeenko_Medvedev_2007,torquato2010jammed,Kapfer_Mickel_Mecke_Schroder-Turk_2012,Mickel_Kapfer_Schroder-Turk_Mecke_2013,klatt2014characterization,klumov2014structural}.
In particular, suitable modifications of bond orientational order parameters have revealed local crystalline order in (non-isostatic) jammed packings~\cite{Kapfer_Mickel_Mecke_Schroder-Turk_2012,Mickel_Kapfer_Schroder-Turk_Mecke_2013}, while icosahedral order is of more limited extent~\cite{klatt2014characterization,klumov2014structural}.
Finally, we can analyse the 
{\it large scale} corresponding to $k \to 0$, 
where fluctuations over the entire sample are considered.
It has been reported 
that systems prepared at the jamming transition 
have unexpected density fluctuations at large scales, possibly corresponding to 
hyperuniform behavior. Physically, hyperuniformity implies that the 
amplitude of volume fraction fluctuations at wave number $k$ vanishes 
as $k \to 0$~\cite{donev2005unexpected,zachary2011hyperuniformity}.
However, several recent works have reported numerical evidence 
that hyperuniformity is not strictly obeyed at the jamming 
transition~\cite{wu2015search,ikeda2015thermal,ikedaparisi}.

The structural properties mentioned above have been studied 
at ``the'' jamming transition. However, theoretical studies 
have established that $\phi_{\rm J}$ is not uniquely determined but 
is in fact strongly protocol dependent~\cite{speedy1998random,torquato2000random,mari2009jamming,biazzo2009theory,hermes2010jamming,chaudhuri2010jamming,ciamarra2010recent,schreck2011tuning,vaagberg2011glassiness,otsuki2012critical,ozawa2012jamming,baranau2014random,baranau2014jamming,bertrand2016protocol,kumar2016memory,luding2016so}.
This implies that the jamming critical point, J-point, 
actually corresponds to a line of critical points, 
thus forming a ``J-line''~\cite{chaudhuri2010jamming}.
A simple way of exploring the J-line is to perform 
compressions of hard sphere configurations using different compression rates.
A more controlled method consists in first thermalising 
the hard sphere fluid at finite temperature at some volume fraction 
and then performing a rapid compression towards 
jamming~\cite{witten09,chaudhuri2010jamming,ozawa2012jamming}.
By varying the volume fraction of the parent fluid, a finite range 
of jamming volume fractions can be explored, while easily keeping
crystallisation under control.
 
Although the existence of a J-line is well accepted, much 
less is known
about how structural properties of jammed systems may change along the J-line.
Some previous studies reported that the structural properties 
at the contact and neighbor scales hardly change~\cite{hermes2010jamming,schreck2011tuning,chaudhuri2010jamming,maiti2014} and thus one might conclude that the structural properties are essentially invariant along the J-line.
Scaling theories of the jamming transition are based on 
this assumption~\cite{van2009jamming,liu2010jamming,goodrich2016scaling}.
On the contrary, some other studies pointed out that there are tiny but 
systematic structural changes~\cite{vaagberg2011glassiness,ozawa2012jamming},
in particular at the neighbor scale.
To our knowledge, the evolution of the large scale 
structural properties along the J-line has not been studied in detail.
Analyzing the evolution of structure 
along the J-line is a numerical challenge, because 
changing the value of $\phij$ requires changing the 
typical preparation timescale over orders of magnitude, 
and even large numerical efforts may yield relatively minute changes
to the value of $\phij$. Therefore, previous studies
have accessed finite, but quantitatively modest, variations of 
$\phij$ along the J-line. 

In this paper, we develop a numerical strategy that allows
us to stretch the extension of the J-line of frictionless hard spheres considerably. 
As a result, we can characterize the variation of structural 
properties with $\phij$ over an unprecedentedly broad range
of volume fractions, 
$\phij \approx 0.65 - 0.70$. This range is at least 3 times wider than 
in any previous study of frictionless jammed packings~\cite{chaudhuri2010jamming,kumar2016memory}. 
The decisive factor allowing the present analysis is the 
recent development of an efficient thermalisation algorithm  
for polydisperse hard sphere fluids. We have recently shown 
that this approach allows the equilibration of very dense 
fluid states~\cite{berthier2016equilibrium}, 
bypassing any alternative method by many orders of magnitude.
Here, we use these deeply 
thermalised fluid configurations as starting point for rapid compressions
towards jammed states.

We find that at the 
{contact scale}, both isostaticity and the same critical behavior 
of the radial distribution function hold over the entire J-line, 
thus corroborating previous
results~\cite{schreck2011tuning,chaudhuri2010jamming}.
To characterize the structure at the {neighbor scale}, 
we apply analysis tools appropriate to polydisperse packings, 
based on the detection of locally favored 
structures~\cite{coslovich2007understanding,Royall_Williams_2015}. Our key finding
is the detection of distorted local icosahedral structures
that become increasingly numerous as $\phij$ is increased, so that
about $80~\%$ of the particles form such structures
in our densest packings. This suggests 
that this local structural motif might be a key geometric factor 
allowing the production of very dense 
jammed packings~\cite{klatt2014characterization}.
At the large scale, 
we find that a nearly hyperuniform 
behavior~\cite{atkinson2016critical} for the smallest
$\phij$, but deviations from hyperuniformity 
increase rapidly as $\phij$ is increased,
suggesting that optimized packings display stronger volume fraction
fluctuations at large scale. 
On the other hand, finite size effects of the mean value of $\phij$ 
are considerably suppressed when $\phij$ is large. 
Overall, our work disentangles contact scale properties
that are deeply connected to the criticality of the jamming
transition, to structural properties at larger scale 
which evolve significantly along the J-line, 
and have therefore a distinct physical origin. 

This paper is organized as follows.
In Sec.~\ref{sec:methods}, we describe the details of the
simulation methods. The resulting extended
range of jamming densities is discussed in  
Sec.~\ref{sec:dramatic}.
We discuss the physics at contact 
scale (isostaticity, pair correlation) in Sec.~\ref{sec:contact},
the physics at the local scale (bond orientational order, 
locally favored structures, rattlers) in Sec.~\ref{sec:local}, 
and at the large scale (hyperuniformity, global finite-size effects)
in Sec.~\ref{sec:global}.
We conclude our paper and discuss our results in Sec.~\ref{sec:conclusions}.

\section{Numerical methods}
\label{sec:methods}

\subsection{The model}

We employ the standard model of additive frictionless hard spheres in three dimensions~\cite{allen1989computer}. 
The pair interaction is zero for non-overlapping particles, infinite otherwise. 
We use a continuous size polydispersity, where the particle diameter $\sigma$ is distributed according to $f(\sigma) = A\sigma^{-3}$, $\sigma \in [\sigma_{\rm min}, \sigma_{\rm max}]$, 
where $A$ is a normalization constant. 
Following previous work~\cite{berthier2016equilibrium},
we use the size polydispersity $\Delta=\sqrt{\overline{\sigma^2} - \overline{\sigma}^2}/\overline{\sigma} = 23 \%$, corresponding to $\sigma_{\rm min} / \sigma_{\rm max} = 0.4492$, where $\overline{\cdots}=\int d \sigma 
f(\sigma)(\cdots)$.
We use $\overline{\sigma}$ as the unit of length.
$f(\sigma)$ used in this study is shown in Fig.~\ref{fig:rattler_analysis}(a).
We simulate systems composed of $N$ particles in a cubic cell with periodic boundary conditions.
We mainly use the system sizes $N=1000$ and $8000$, but we also 
perform selected simulations for $N=150$, $300$, $600$, $2000$, $4000$ 
to systematically investigate the finite-size effects
described in Sec.~\ref{sec:global}.
The state of the hard sphere system is uniquely 
characterized by the volume fraction 
$\phi=\pi N \overline{\sigma^3}/(6V)$, where $V$ is the volume of the system.
For the fluid state, we measure the reduced pressure $p=P/(\rho k_{\rm B}T)$, where $\rho=N/V$, $k_{\rm B}$ and $T$ are the number density, Boltzmann constant and temperature, respectively.
We set $k_{\rm B}$ and $T$ to unity.
The pressure $P$ is calculated from the contact value of the pair correlation function properly scaled for a polydisperse system~\cite{santos2005contact}.
The fluid state has a finite $p$, whereas the jammed state corresponds to $p \to \infty$.

Note that the functional form of $f(\sigma)$ with $\Delta=23 \%$ is chosen so that the system is fairly robust against crystallization and fractionation at extremely high densities~\cite{ninarello2017}.
With smaller $\Delta$, the system easily crystallizes when using the efficient swap Monte Carlo method described below.
On the other hand, as demonstrated in previous studies, large values of $\Delta$ or different forms of $f(\sigma)$ lead to fractionation at sufficiently high density~\cite{wilding2004phase,wilding2010phase}.
By contrast, the model parameters employed in this work are optimized to avoid such instabilities and thus enable us to explore the J-line over an unprecedented range of packing fractions.

\subsection{Equilibration of very dense fluid states}

To obtain equilibrium fluid configurations, we perform Monte Carlo (MC) simulations which combine 
traditional translational particle displacements and 
non-local particle swaps~\cite{sindzingre1989calculation,gazzillo1989equation,santen2001liquid,grigera2001fast,brumer2004numerical,fernandez2007phase,gutierrez2015static,ninarello2017}.
Translational displacements are drawn from a cube of linear side $0.115$,
and a trial displacement is accepted if it does not 
create an overlap between particles~\cite{allen1989computer}.  
For a trial swap move, we randomly pick a pair $(i,j)$ of particles 
with $|\sigma_i - \sigma_j|< a$
(we choose $a=0.2$) and attempt to exchange their diameters~\cite{ninarello2017}. 
The swap is accepted if it does not create an overlap.
We perform translational moves with probability $0.8$, 
and swap moves with probability 0.2~\cite{ninarello2017}. 
We have previously established~\cite{berthier2016equilibrium}
that this swap Monte Carlo 
setting is extremely efficient to thermalise hard sphere 
fluid states up to very large volume fractions,
$\phi_{\rm fluid} \approx 0.655$. We have checked
that all our equilibrium configurations are
taken in the fluid state, carefully monitoring possible signs 
of demixing or crystallization using the same structural 
tools described below to also characterize jammed states.
Therefore, the parent fluid configurations used to produce jammed states
all belong to the (metastable) equilibrium fluid branch, 
and we do not artificially vary the jamming density by introducing 
partially crystallized or demixed states. Instead, we explore the equilibrium
fluid branch over a broad range of densities.

\subsection{Compression towards the jamming transition}
\label{sec:compression}

Having prepared equilibrium fluid configurations of hard spheres,
we use the non-equilibrium compression algorithm introduced
in Refs.~\cite{xu2005random,desmond2009random} to reach the 
jamming transition.
Briefly, this algorithm replaces the hard sphere potential with a soft 
repulsive harmonic potential, given by
\begin{equation}
v_{ij}(r_{ij})=\frac{\epsilon}{2} \left[ 1- \left( r_{ij}/\sigma_{ij} \right) \right]^2 \theta (1-r_{ij}/\sigma_{ij}),
\label{eq:harmonic}
\end{equation}
where $\epsilon$ is the energy scale, $r_{ij}$ is the distance between 
the spheres $i$ and $j$,
$\sigma_{ij}=(\sigma_i + \sigma_j)/2$ 
is the distance of the two spheres at contact,
and $\theta(x)$ is the Heaviside step function.
We use $\epsilon$ as the unit of energy.
The idea of the algorithm is to 
alternate instantaneous compression steps and 
energy minimization to iteratively converge to the jamming density. 

During a compression step, we 
increase the volume fraction of the system $\phi$ by 
$\delta \phi = 5 \times 10^{-4}$.
This compression introduces finite overlaps between particles, such 
that the  potential energy $U = \sum_{i<j} v_{ij}(r_{ij})$ becomes 
finite.  
These overlaps are then eliminated by performing an energy minimization.
To this end, we apply the conjugate gradient minimization 
method~\cite{nocedal1999numerical} for $U$.

If the system still has a finite $U$ after minimization, 
we decrease $\delta \phi$ by factor of $2$ and use a series of
decompression and compression steps until the overlaps are eliminated.
This process is interpreted as pulling the system from trivial local 
potential energy minima.
We stop the algorithm when $\delta \phi < 1.0 \times 10^{-6}$,
and the resulting system is essentially a hard sphere jammed packing.
However, we find that the algorithm explained so far~\cite{desmond2009random} produces a fraction of slightly hypostatic packings.
Thus, we perform an additional process introduced in Ref.~\cite{chaudhuri2010jamming} to get more accurate isostatic 
jammed packings.
We simulate sequential compression and minimization with $\delta \phi = 1.0 \times 10^{-5}$ until $U/N  > 1.0 \times 10^{-6}$.
Then, we decompress the system with $\delta \phi=1.0 \times 10^{-6}$ until $U/N < 10^{-16}$, which determine the jamming transition 
point~\cite{chaudhuri2010jamming}.

Note that the jamming algorithm employed in this work is based on 
purely isotropic compressions and decompressions.
Thus, stability against shear deformation is not 
guaranteed~\cite{dagois2012soft}, but we expect that further 
optimization against shear would not change our conclusions.

\section{Extending the range of jamming densities}

\label{sec:dramatic}

\begin{figure}
\begin{center}
\includegraphics[width=1.\columnwidth]{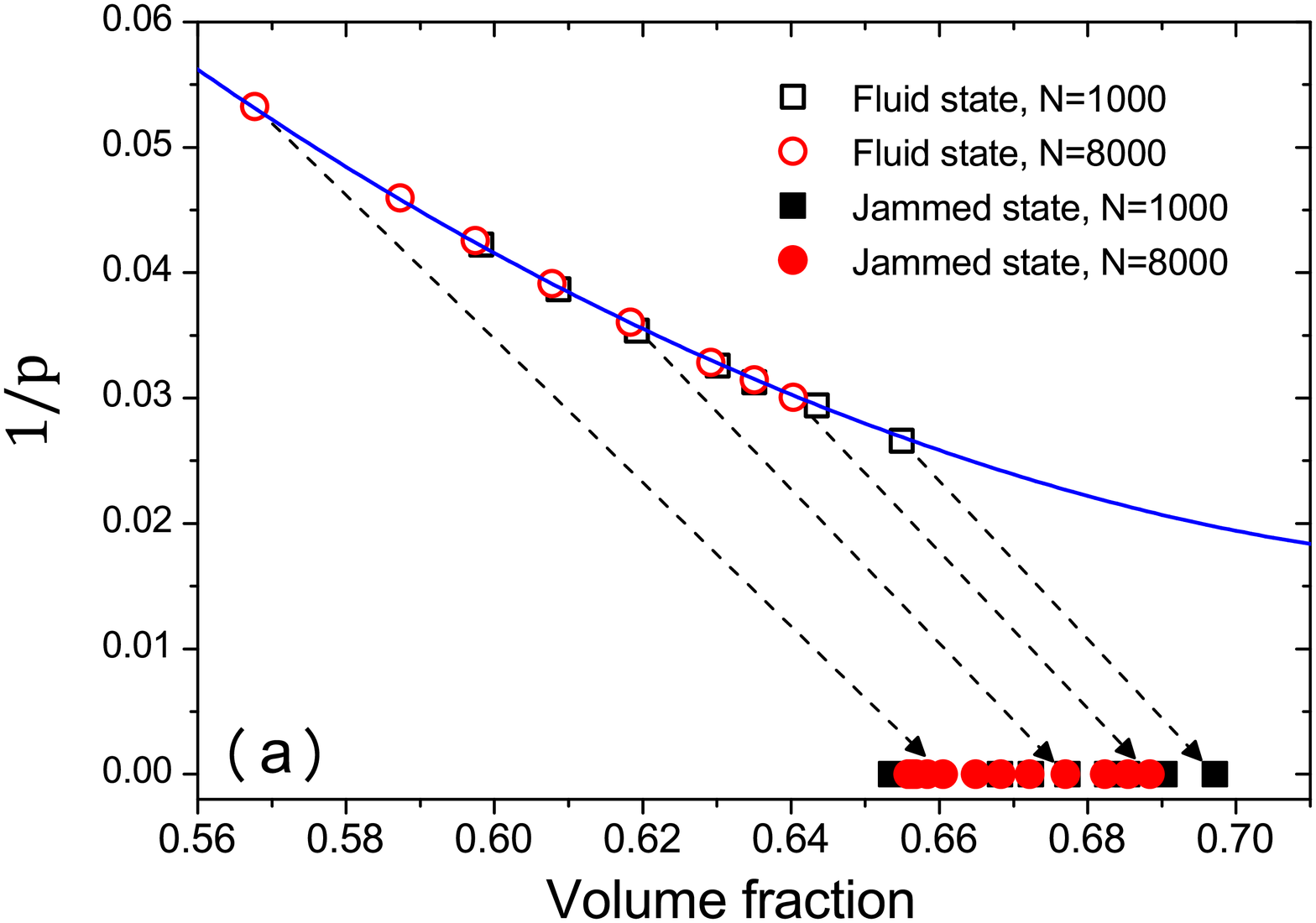}
\includegraphics[width=1.\columnwidth]{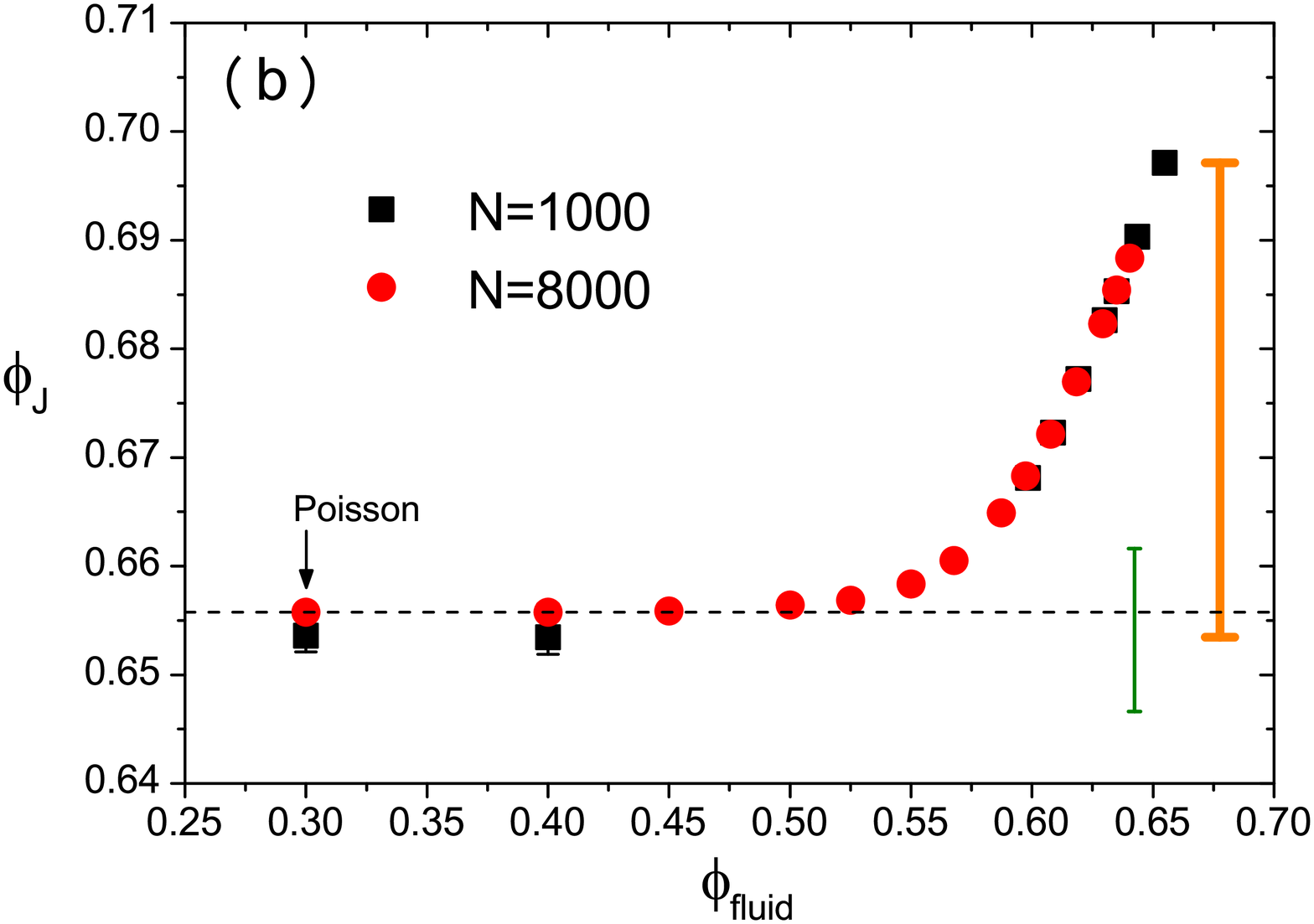}
\caption{
(a) Equation of state of the system for two different system sizes. 
Equilibrium fluid states are shown as empty points.
A modified version of the Carnahan-Starling equation of state for polydisperse systems is drawn as a solid line~\cite{berthier2015growing}. 
Jammed states (filled points) obtained by the compression algorithm are located at $p \to \infty$.
The dashed arrows connect an equilibrium parent fluid state to the corresponding jammed state.
(b) Evolution of the averaged jamming transition volume 
fraction $\phi_{\rm J}$ obtained by the compression from an equilibrium parent fluid with volume fraction $\phi_{\rm fluid}$. We also report 
$\phi_{\rm J}$ starting from a Poisson distributed configuration with
$\phi_{\rm fluid}=0.3$. The horizontal dashed line is $\phi_{\rm J}$ generated from a Poisson distribution for $N=8000$.
The vertical bars correspond to the width of the 
J-line for this work (thick bar) which is about 3 times 
larger than previous works using 
a binary mixture~\cite{chaudhuri2010jamming} (thin bar).}
\label{fig:phi_vs_phiJ}
\end{center}
\end{figure}

We present the equation of state of the system in Fig.~\ref{fig:phi_vs_phiJ}(a).
Equilibrium fluid configurations at finite pressure $p$ are compressed using the non-equilibrium algorithm described in Sec.~\ref{sec:compression} towards jammed states at $p \to \infty$.
Figure~\ref{fig:phi_vs_phiJ}(a) demonstrates that the higher the volume fraction of the parent fluid, $\phi_{\rm fluid}$, the higher the jamming transition volume fraction, $\phi_{\rm J}$.
Thus, enhanced thermalization is the key to extend the J-line.

In Fig.~\ref{fig:phi_vs_phiJ}(b), we show $\phi_{\rm J}$  as a function of $\phi_{\rm fluid}$, varying $\phi_{\rm fluid}$ over a very broad range.
Below $\phi_{\rm fluid} \lesssim 0.53$, the observed $\phi_{\rm J}$ is almost 
independent of $\phi_{\rm fluid}$, as suggested by our horizontal
dashed line. 
Also, we show the result obtained for a compression 
starting from a Poisson distributed system of harmonic soft spheres 
with $\phi_{\rm fluid}=0.3$. We find that 
$\phi_{\rm J}$ from the Poisson distribution takes the same value as 
from dilute hard sphere fluid configurations, $\phi_{\rm J} \simeq 0.655$.
Thus, we confirm that the protocol dependence of $\phi_{\rm J}$ is essentially absent at $\phi_{\rm fluid} \lesssim 0.53$.
This value for $\phij$ 
is consistent with recent numerical results for a similar continuously 
polydisperse system~\cite{desmond2014influence}.

As expected, $\phi_{\rm J}$ starts to depart from the plateau value $\phi_{\rm J} \simeq 0.655$ when $\phi_{\rm fluid} \gtrsim 0.53$, and it then monotonically 
increases with increasing $\phi_{\rm fluid}$. The largest value we 
obtain is $\phi_{\rm J} \simeq 0.7$.
A qualitatively similar behavior was observed before 
in a binary hard sphere mixture~\cite{chaudhuri2010jamming,ozawa2012jamming}.
The variation of the jamming density with the initial fluid density 
is reminiscent of the variation of the energy of inherent structures
with initial temperature for glass-forming 
materials with a continuous pair 
interaction~\cite{sastry1998signatures,brumer2004mean,ozawa2012jamming}.
Note that the finite size effect of $\phi_{\rm J}$ between $N=1000$ 
and $8000$ is small at low $\phi_{\rm fluid}$, and it essentially 
vanishes in the large $\phi_{\rm fluid}$ regime, at least in this 
graphical representation.
The finite size effect of $\phi_{\rm J}$ will be more systematically examined 
in Sec.~\ref{sec:global}.
We draw as vertical bars the widths of the J-line obtained in the 
present study (thick bar) and in a previous study of a binary 
mixture~\cite{chaudhuri2010jamming} (thin bar).
The extension of the J-line achieved in this work is about $3$ times 
larger than previous works~\cite{chaudhuri2010jamming,kumar2016memory}.
Thus, we have indeed succeeded in stretching the J-line considerably.
In the next sections, we analyze the structural properties 
of the obtained packings along this stretched J-line.

\section{Structure at the contact scale: Isostaticity
and pair correlation}
\label{sec:contact}

\subsection{Isostaticity}

We first examine structural properties at the {contact scale}.
The relevant length scale for $\phi \le \phij$
is the gap between particles, $x-1$, 
where $x=r_{ij}/\sigma_{ij}$.
A central quantity describing jammed states at this contact scale 
is the averaged contact number, $Z$.
In practice, two particles, $i$ and $j$, are considered in contact 
if $r_{ij} \leq (1+a)\sigma_{ij}$, where $a=1 \times10^{-5}$ and $5 \times 10^{-6}$ for $N=1000$ and $8000$, respectively~\cite{schreck2011tuning}.
With this definition, 
$Z$ is given by $Z=N_{\rm c}/(N-N_{\rm r})$, where $N_{\rm r}$ is the number of the rattlers, $N_{\rm c}$ is the number of contact pairs among $N-N_{\rm r}$ particles which make a contact network.
We define a rattler particle as a particle whose contact number is 
smaller than $d+1$, where $d$ is the spatial dimension.
Note that a determination of rattlers has to be performed 
iteratively for a given configuration until all rattlers are removed.
The isostatic packing is defined by the condition $Z=2d$ (thus, $Z=6$
in three dimensions), reflecting the fact that the system is mechanically
marginally stable~\cite{muller2015marginal}. 

\begin{figure}
\begin{center}
\includegraphics[width=0.95\columnwidth]{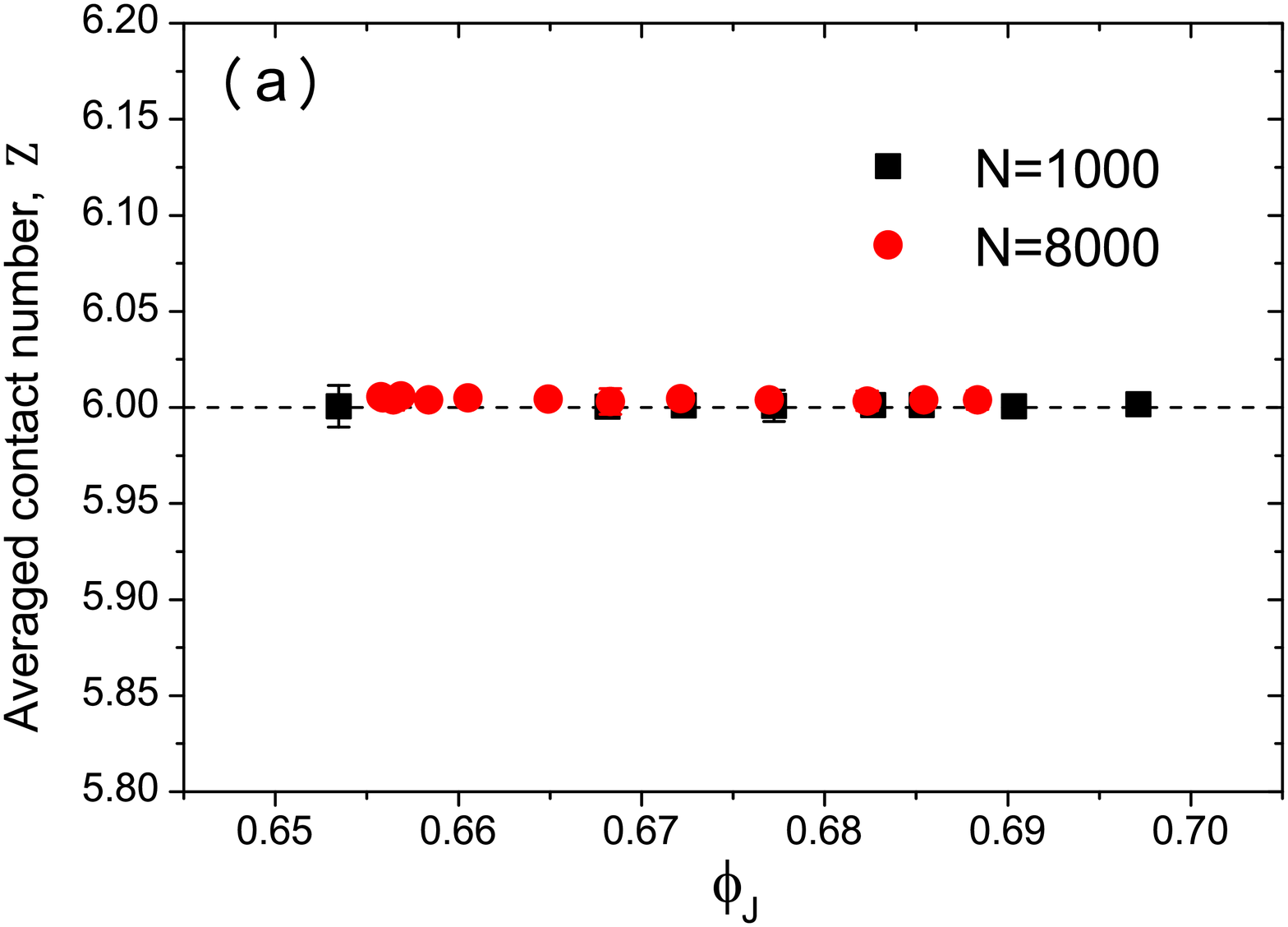}
\includegraphics[width=0.95\columnwidth]{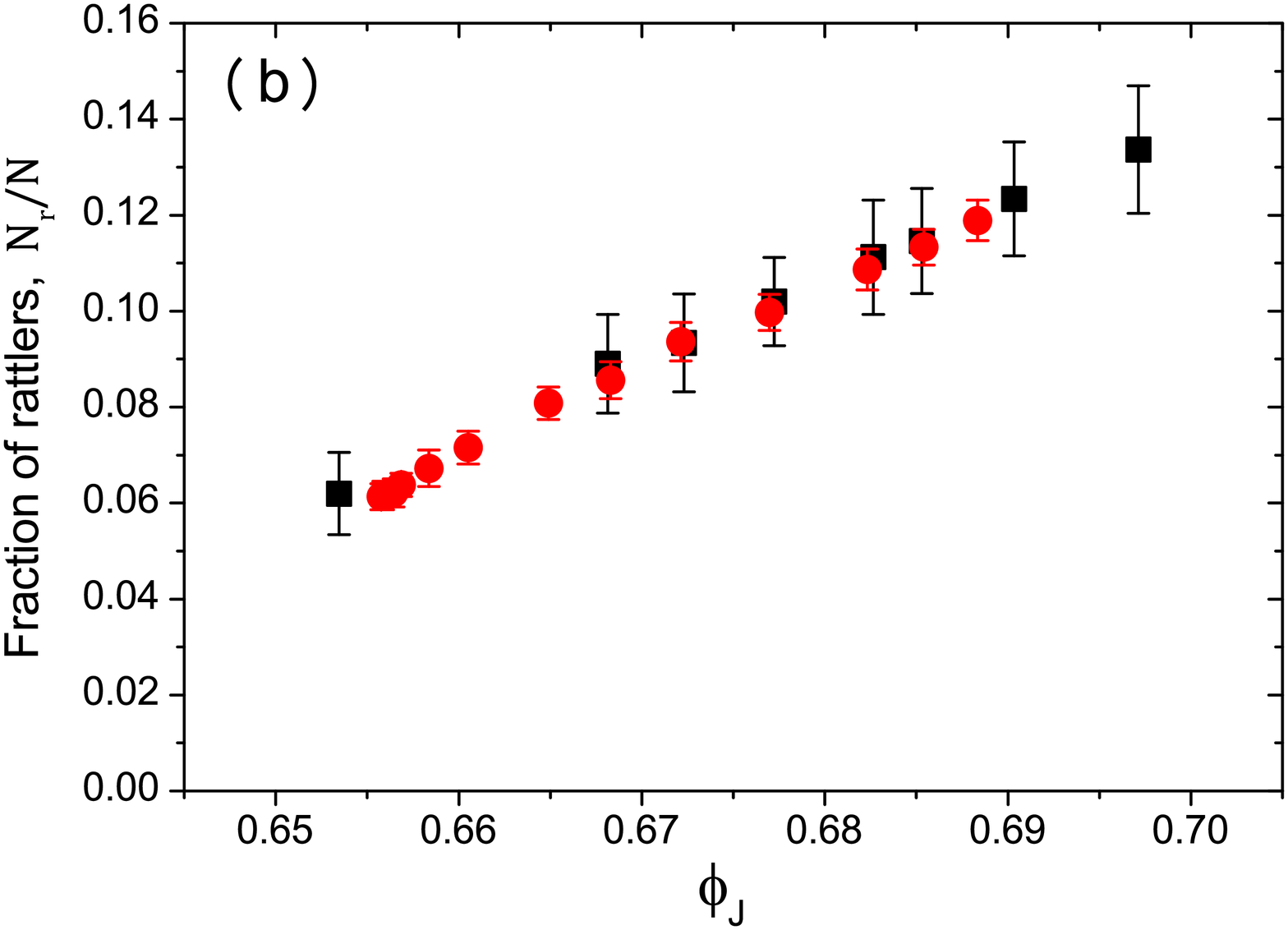}
\caption{
(a): The averaged contact number $Z$ as a function of $\phi_{\rm J}$
remains close to the isostatic value $Z=6$ along the
entire J-line. 
(b): The fraction of rattlers, $N_{\rm r}/N$, 
increases substantially as a function of $\phi_{\rm J}$.}
\label{fig:contact_number}
\end{center}
\end{figure} 

In Fig.~\ref{fig:contact_number}(a), we show $Z$ as a function of $\phi_{\rm J}$
over the entire range of jamming densities that we were able to explore.
We find that the isostatic condition, $Z=6$, indeed holds over
the entire J-line obtained by our protocols.
In Fig.~\ref{fig:contact_number}(b) we show 
the fraction of rattlers, $N_{\rm r}/N$.
It is around $6 \%$ at the lowest jamming density, 
$\phi_{\rm J} \simeq 0.655$, which is comparable with previous results 
using energy minimization in three dimensions for 
binary~\cite{o2003jamming,chaudhuri2010jamming} and 
polydisperse~\cite{zhang2015structure} systems.
Interestingly, the fraction of rattlers increases steadily with $\phi_{\rm J}$.
Note that a slight increase of the fraction of rattlers along the J-line 
was also reported previously in a 
binary mixture~\cite{chaudhuri2010jamming,maiti2014}.
This growth of the number of rattlers might appear counterintuitive 
at first sight, because rattlers tend to occupy a larger volume~\cite{maiti2014}
and increasing their number should decrease the efficiency of the packing,
in contradiction with the results shown in Fig.~\ref{fig:contact_number}(b).
The issue of the increase of the number of 
rattlers will be further discussed in Sec.~\ref{sec:rattlers}.
 
\subsection{Pair correlation function at contact}

\begin{figure}
\begin{center}
\includegraphics[width=0.95\columnwidth]{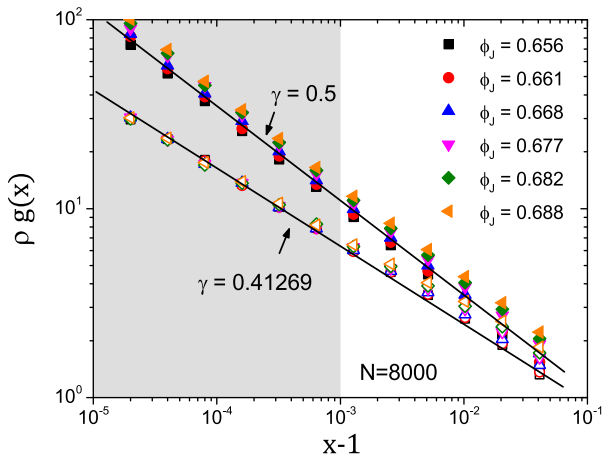}
\caption{Radial distribution function $\rho g(x)$ near contact, 
where $\rho=N/V$ and $x=r_{ij}/\sigma_{ij}$.
The closed and open symbols are obtained with and without ratters, respectively.
The black straight lines characterize a power law behavior, 
$g(x) \propto (x-1)^{-\gamma}$, with $\gamma = 0.5$ (the upper line) 
and $\gamma = 0.41269$ (the lower line), respectively.
When rattlers are removed, a universal power law with the value predicted 
by the mean-field theory~\cite{charbonneau2014fractal} is observed for all $\phij$ values.
The gray shaded region marks the regime, $x-1 \lesssim 10^{-3}$, where critical scaling with $\gamma = 0.41269$ holds. 
}
\label{fig:gofx_critical}
\end{center}
\end{figure}  

It is known that the radial distribution function shows a power law critical behavior near contact, $g(x) \propto (x-1)^{-\gamma}$, for isostatic packings.
We show $\rho g(x)$, where $\rho=N/V$, as a function of the gap, $x-1$, in Fig.~\ref{fig:gofx_critical} for several $\phi_{\rm J}$ values along the J-line.
Note that we multiply $g(x)$ by $\rho$ to remove the trivial effect 
of the density change. 
The data for all $\phi_{\rm J}$'s follow a power law with an exponent
$\gamma=0.5$, when we compute $g(x)$ from all particles 
in a given configuration, in particular including rattlers~\cite{silbert2006structural}.
Instead, when the rattlers are removed and only the particles participating in the contact network are taken into account~\cite{lerner2013low}, the $g(x)$ for all $\phi_{\rm J}$'s now follow a distinct power law which is 
compatible with the value $\gamma=0.41269$
predicted by the mean-field theory of the jamming 
transition~\cite{charbonneau2014fractal}.
Thus, a careful treatment of the rattlers is essential to assess 
the critical behavior of the pair correlation function at 
contact~\cite{lerner2013low,charbonneau2012universal}.

From these observations, we are able to confirm that the critical 
behavior in $g(x)$ holds over the entire J-line and is therefore universal.
This is not a trivial observation. For instance, it may have happened 
that this property
only holds at the lowest end of the J-line, which is the only 
point where jamming criticality and mean-field predictions had been analyzed 
before~\cite{charbonneau2015jamming}. Notice also 
that establishing this result was not easy. We found appreciable deviations 
from a power law behavior for smaller systems at low
values of the argument $x-1$ (not shown). On the other hand, we also find that 
by increasing $\phij$ the power law regime is entered for lower 
values of $x-1$, see Fig.~\ref{fig:gofx_critical}. Therefore, observing 
a power law for large $\phij$ is difficult, as the critical regime 
becomes less easily observed for a given system size as $\phij$ is increased. 

We expect that other critical behaviors such as a power law 
distribution of the contact forces are also observed over the entire J-line 
since these critical behaviors are all directly connected to one 
another~\cite{charbonneau2014fractal,muller2015marginal}.
However, this direction is beyond the scope of this paper, since it would 
require a more important computational 
effort to prepare the packings exactly at the 
jamming transition~\cite{charbonneau2015jamming}. 

\section{Structure at the local scale}
\label{sec:local}

Our results so far confirm that the combined use of the swap Monte Carlo technique and polydisperse spheres stretch 
substantially the J-line, while maintaining isostaticity, $Z=6$.
This already indicates that the system does not have crystalline order, 
which would produce $Z>6$ (hyperstatic packings),  
as explicitly demonstrated in 
Ref.~\cite{schreck2011tuning}.
For monodisperse spheres, packing fractions comparable and even larger to the ones observed here can be attained. 
However, these packings are not isostatic and there is evidence that they accumulate local crystalline order beyond a threshold density~\cite{Kapfer_Mickel_Mecke_Schroder-Turk_2012}.
Thus, the next question is: How do jammed configurations of polydisperse spheres attain such large densities while remaining amorphous?
In this section, we address this issue by investigating the subtle evolution of the geometry of local packing at the {neighbor scale}, $k \sim 2 \pi/\overline{\sigma}$.
In particular, we confirm a mild increase of bond orientational order upon increasing density (Sec.~\ref{sec:boo})
and reveal the emergence of distorted icosahedral local structures (Sec.~\ref{sec:lfs}).
Above the onset density, these structural features are present in both fluid and jammed states.

\subsection{Bond-orientational order}\label{sec:boo}

\begin{figure*}
\begin{center}
\includegraphics[width=0.95\columnwidth]{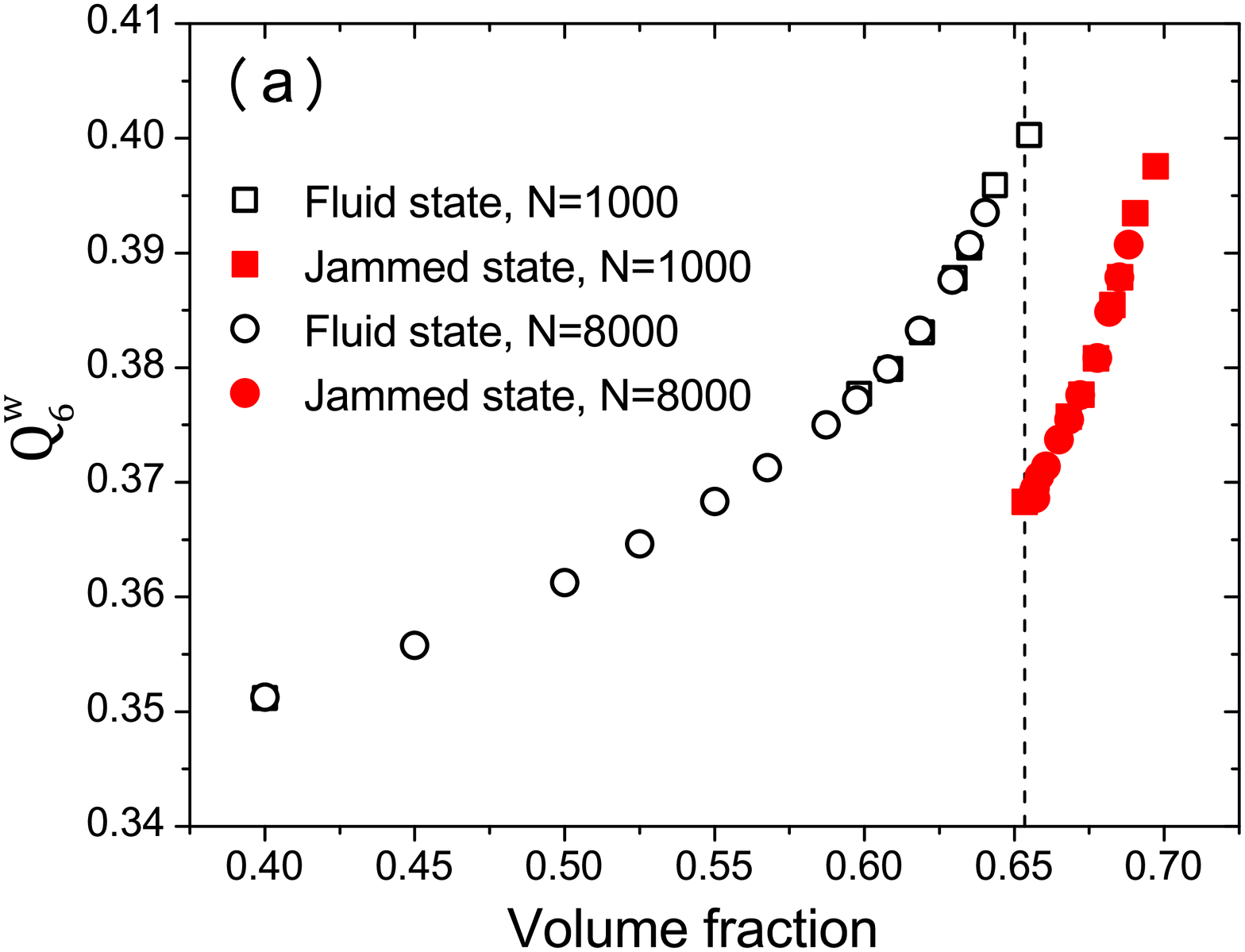}
\includegraphics[width=0.95\columnwidth]{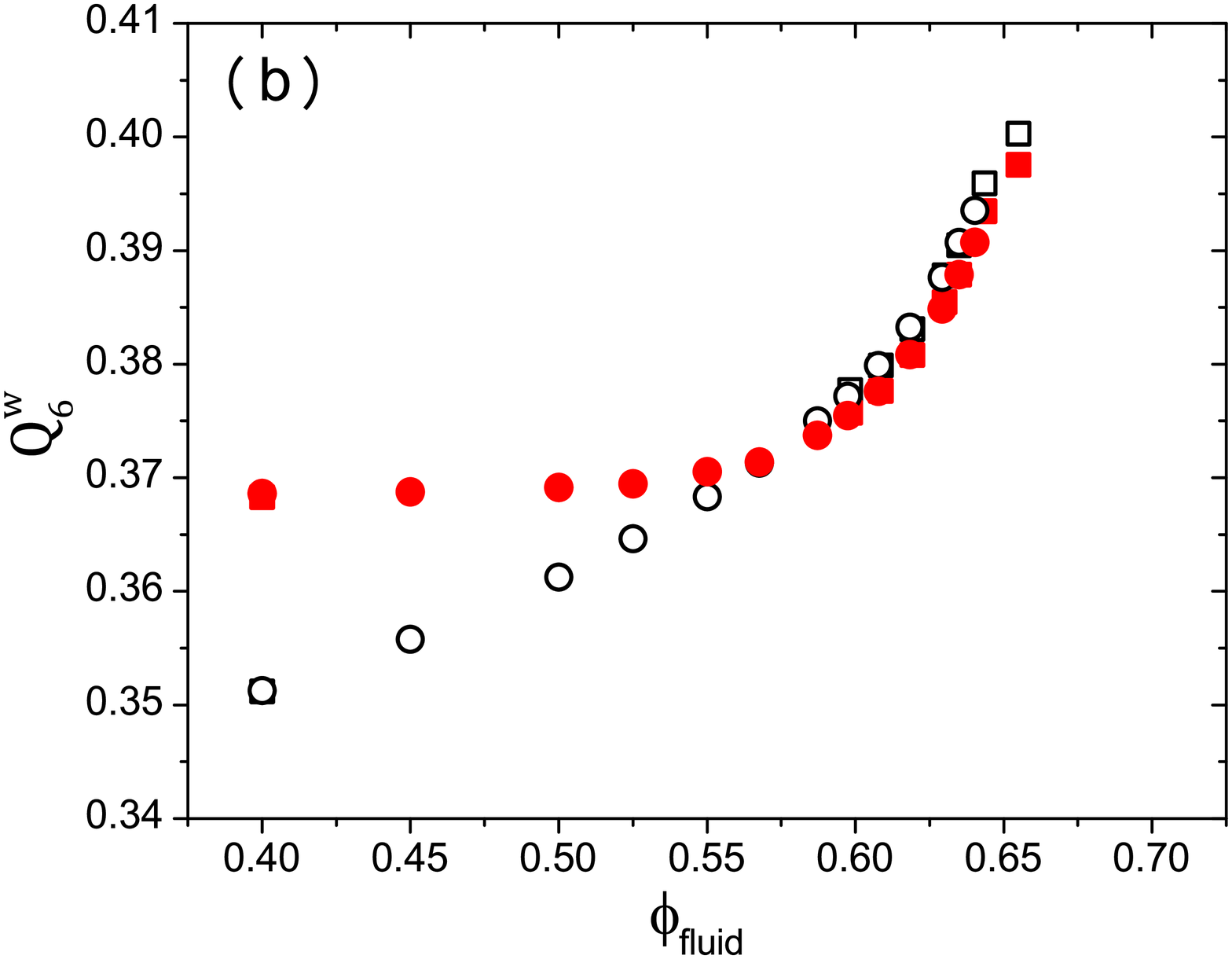}
\includegraphics[width=0.95\columnwidth]{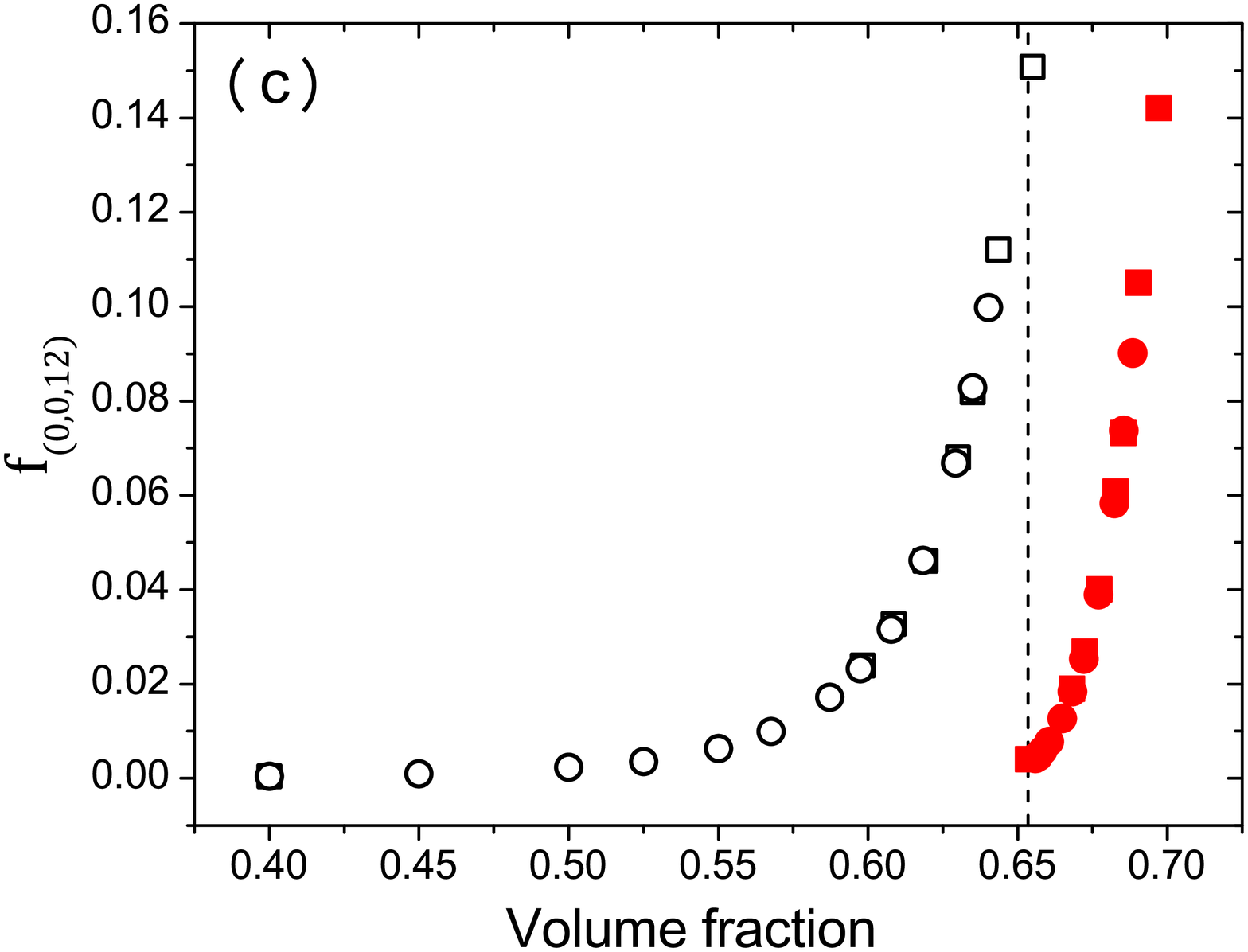}
\includegraphics[width=0.95\columnwidth]{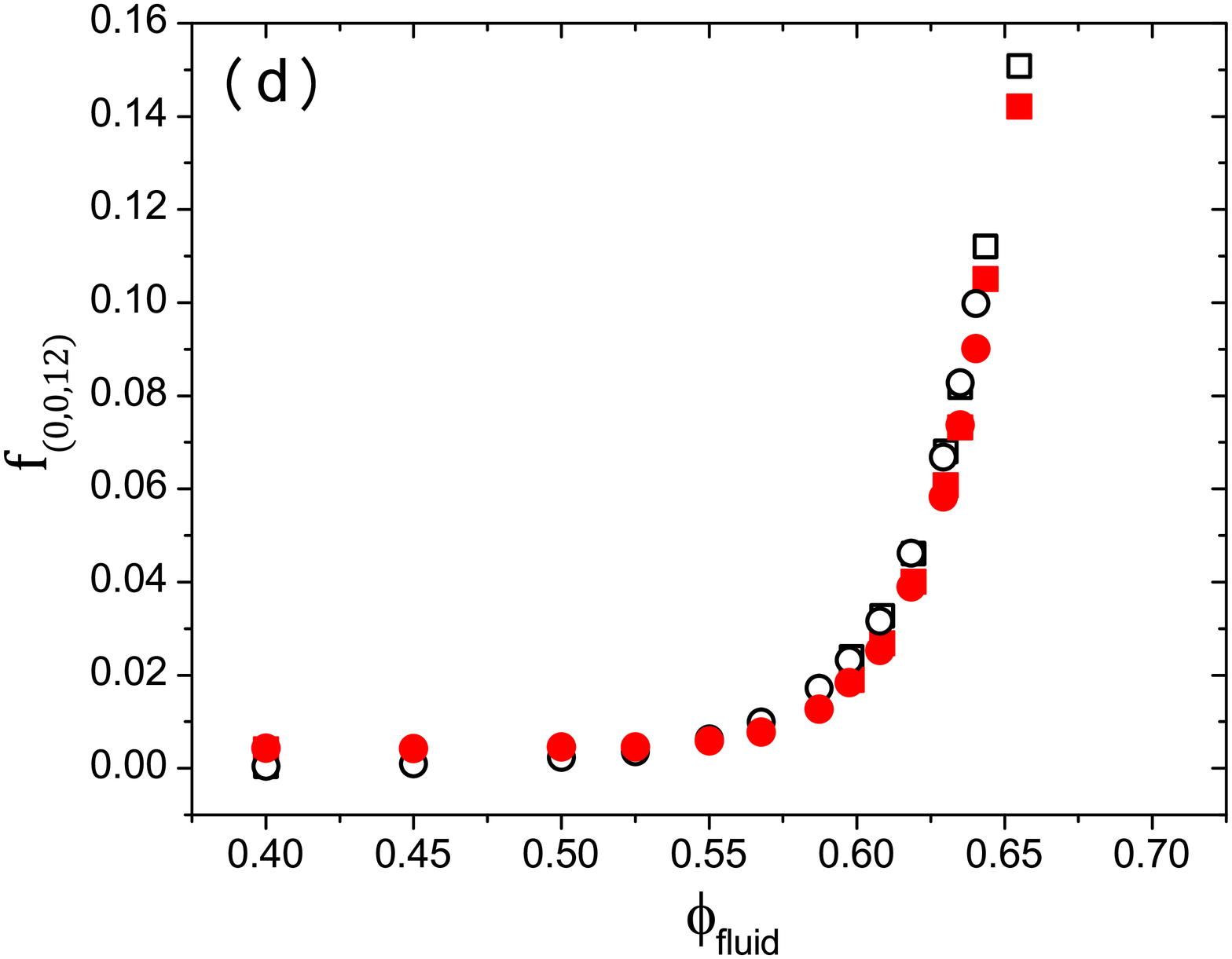}
\caption{
(a, b): Bond orientational order parameter $Q_6^w$ for the fluid and 
jammed states as a function of their volume fraction (a), 
or the volume fraction of the parent fluid, $\phi_{\rm fluid}$, (b).
(c, d): Similar plots for the fraction of icosahedral structures characterized by the $(0,0,12)$ Voronoi signature.
The vertical lines are a guide for the eye.}
\label{fig:Q_and_LFS}
\end{center}
\end{figure*}  

Bond-orientational order (BOO) parameters~\cite{steinhardt1983bond} have been used extensively to characterize the local structure of jammed packings~\cite{torquato2000random,ozawa2012jamming,klumov2014structural}.
The idea is to compute rotational invariants from a multiple expansion of the nearest neighbors distance distribution, and compare the results to the values observed for known crystal structures~\cite{steinhardt1983bond}.
Inspection of the combined distribution of BOO parameters of different orders allows one to disentangle structures characterized by different local symmetries~\cite{Moroni_2005}.
For a given particle $i$, the local BOO parameter is defined as
\begin{equation}
Q_{l,i}=\sqrt{ \frac{4 \pi}{2l + 1} \sum_{m=-l}^{l} \left| \frac{1}{n_{\rm b}(i)} \sum_{j=1}^{n_{\rm b}(i)} Y_{l,m}({\bf r}_{ij}) \right|^2 },
\end{equation}
where $n_{\rm b}(i)$ is the number of nearest neighbors of the $i$-th particle, $Y_{l,m}({\bf r}_{ij})$ is the spherical harmonic of degree $l$ and order $m$, and ${\bf r}_{ij}$ is the vector distance between particle $i$ and $j$.
The average BOO parameters, $Q_{l} = (1/N) \sum_i Q_{l,i}$, can then be used to characterize the local structure of a packing.

It is well known that the values of local BOO parameters in dense, disordered packings depend sensitively on the definition of the neighbors network surrounding each particle~\cite{Mickel_Kapfer_Schroder-Turk_Mecke_2013}.
In disordered packings, neighbors are often defined as particles whose distance contributes to the first peak of the radial distribution function; alternatively, the nearest neighbors network is obtained from a Voronoi tessellation of the particles' coordinates~\cite{Gellatly_Finney_1982}.
In order to cure the sensitivity of BOO parameters to the details of neighbors network, Mickel \textit{et al.}~\cite{Mickel_Kapfer_Schroder-Turk_Mecke_2013} have proposed to weight each bond entering the calculation of $Q_{l,i}$ by the area of the corresponding face of the surrounding Voronoi cell.
The resulting weighted BOO parameters are then defined as
\begin{equation}
Q_{l,i}^w=\sqrt{ \frac{4 \pi}{2l + 1} \sum_{m=-l}^{l} \left| \sum_{j=1}^{n_{\rm b}(i)} \frac{A_{ij}}{A_i} Y_{l,m}({\bf r}_{ij}) \right|^2 },
\end{equation}
where $A_{ij}$ denotes the area of the face of Voronoi cell connecting particle $i$ and $j$ and $A_i = \sum_j A_{ij}$.
A similar idea had been suggested long ago in Ref.~\cite{steinhardt1983bond}, to cure the artifacts mentioned above. %and the numerical instability inherent in Voronoi tessellations.
In the following, we compute the weighted BOO employing the neighbors network determined via a radical Voronoi tessellation~\cite{Gellatly_Finney_1982} obtained using Voro++~\cite{voro++}.

A standard indicator of structural ordering is the average $Q_6$, which takes large values for ordered, close-packed structures ($Q_6=Q_6^w=0.575$ for the fcc cell and $Q_6=Q_6^w=0.663$ for the icosahedron).
In Figs.~\ref{fig:Q_and_LFS}(a, b), we show the weighted BOO parameter $Q_6^w$ for both equilibrium fluid (open symbols) and jammed (closed symbols) states.
The data are shown as a function of the corresponding volume fraction of jammed states, $\phi_{\rm J}$, and as a function 
of the volume fraction of the parent equilibrium fluid, ($\phi_{\rm fluid}$), in panels (a) and (b), respectively.
In both fluid and jammed states, $Q_{6}^w$ increases monotonically with increasing packing fraction, 
indicating a systematic, progressive local ordering along the J-line.
In agreement with the results of Ref.~\cite{Mickel_Kapfer_Schroder-Turk_Mecke_2013}, we found that the values of the standard $Q_6$, obtained using the Voronoi neighbors network are smaller than for its weighted counterpart, $Q_6^w$ (by 10-15\% in our case, result not shown).

The different representations of the data in Fig.~\ref{fig:Q_and_LFS} convey two distinct messages: Panel (a) shows that fluid and jammed states can have a similar volume fraction and yet clearly different local structure, as illustrated by the vertical dashed line around $\phi_{\rm J}=0.655$~\cite{yamchi2015inherent,berthier2016equilibrium}.
Panel (b) 
demonstrates instead that above the onset volume fraction, $\phi_{\rm onset} \simeq 0.56$, around which glassy dynamics starts to manifest in the equilibrium fluid~\cite{berthier2017breaking}, the equilibrium fluid and the corresponding jammed states are characterized by very similar bond orientational order; at the local scale, jammed configurations retain essentially the local structure of the parent fluid.
This result is confirmed by analysis of the 
other local structure metrics detailed below.

The evolution of $Q_{6}^w$ is generally not enough to detect the presence of local crystal structures in dense packings.
In particular, it is difficult to disentangle local fcc structures from distorted icosahedra~\cite{Mickel_Kapfer_Schroder-Turk_Mecke_2013}, which both give similar $Q_6$ values.
To resolve fine details of the local structure, one has to resort to scatter plots of rotational invariants of different orders~\cite{steinhardt1983bond}, and possibly introduce an additional averaging 
procedure over the neighbors~\cite{lechner_accurate_2008}.
Mickel \textit{et al.}~\cite{Mickel_Kapfer_Schroder-Turk_Mecke_2013} have suggested to measure $Q_2^w$ invariants, which are zero for most crystal structures 
as a simpler approach to quantify the degree of local crystalline order.
We found that $Q_2^w$ decreases mononotically by increasing volume fraction for both fluid and jammed configurations, and that its variation is smooth and continuous (not shown).
Typical values of the average $Q_2^w$ range from about 0.1 at $\phi=0.4$ to 0.05 at the largest packing fractions.
These values are consistent with those observed in non-crystalline packings of monodisperse hard spheres~\cite{Mickel_Kapfer_Schroder-Turk_Mecke_2013}.
This, together with the smooth evolution of the fluid equation of state and the regularity of the partial structure factors at small $k$~\cite{berthier2017breaking}, confirms that our packings are fully amorphous along the whole J-line.

\subsection{Locally favored structure}\label{sec:lfs}

A slight but systematic change of bond orientational order along the J-line has been noted before~\cite{vaagberg2011glassiness,ozawa2012jamming}, albeit over a smaller range of volume fractions.
While this suggests some sort of ordering of the packings, uncovering the precise nature of local order remains difficult and requires knowledge of a dictionary of ``known'' reference structures.
In this section, we employ a strategy based on the statistics of Voronoi cell shapes~\cite{Finney_1976,Tanemura_Hiwatari_Matsuda_Ogawa_Ogita_Ueda_1977} which does not rely on any a priori reference to specific motifs.
This approach has been employed successfully in the context of glasses, see e.g. Refs.~\cite{Ma_2015,Royall_Williams_2015} for recent reviews, but its use in the context of jammed packings remains very
limited~\cite{klatt2014characterization}.

As in Sec.~\ref{sec:boo}, we perform a radical Voronoi tessellation~\cite{Gellatly_Finney_1982} of each configuration.
In this construction, which duly accounts for the size polydispersity, each particle $i$ in the system is enclosed in a polyhedral cell such that all points whose tangent distance from the surface of particle $i$ is smaller than the tangent distance from the surface of particle $j$ with $j \ne i$~\cite{Gellatly_Finney_1982}.
The shape of a Voronoi cell 
encodes detailed information about the local arrangements of the neighbors around the central particle, hence the local structure.
We characterize the shape of a cell through its signature $(n_3,n_4,n_5,...)$, where $n_q$ is the number of faces of the cell with a given number $q$ of vertices~\footnote{We disregard null values of $n_q$ for $q>q^\prime$, where $q^\prime$ is the largest number of vertices such that $n_q^\prime>0$.}.
By averaging over an ensemble of configurations, we detect the most frequent signatures and monitor their concentration as a function of the relevant control parameter (here the volume fraction).
This approach provides a fairly robust assessment of the preferred local order, often termed ``locally favored structure'' in simple models of glasses~\cite{Royall_Williams_2015}.

We find that at low volume fractions the fluid exhibits a variety of different Voronoi signatures, typical of highly disordered fluids.
Upon increasing $\phi_\textrm{fluid}$, however, fluid samples become increasingly rich in $(0,0,12)$ signatures, which are associated to icosahedral local order.
This kind of signature is by far the most frequent one at largest volume fractions.
In Figs.~\ref{fig:Q_and_LFS}(c, d), we show the fraction $f_{(0,0,12)}$ of icosahedral structures for both fluid and jammed states. As for panels (a) and (b), we plot the data as a function of $\phi_{\rm J}$ and of $\phi_{\rm fluid}$.
We find that $f_{(0,0,12)}$ for both fluid and jammed states is essentially negligible below $\phi_{\rm fluid} \simeq 0.56$, but then it increases rapidly with increasing volume fraction to reach a value of about $14\%$ for our densest packings. 
We note that if we also take into account all particles connected to centers of icosahedral structures, as done for instance in~\cite{malins_identification_2013}, this fraction reaches about 80~\% at the largest density.
Superficially, this behavior resembles the one found in the binary Lennard-Jones mixture introduced in Ref.~\cite{wahnstrom}, for which $f_{(0,0,12)}$ shows a marked increase around the onset of slow dynamics~\cite{coslovich2011locally}.

\setlength{\tabcolsep}{5pt}
\begin{table}
\begin{center}
\begin{tabular}{llllll}
\hline
\hline
\multicolumn{2}{c}{$\phi_{\rm J}=0.656$} &
\multicolumn{2}{c}{$\phi_{\rm J}=0.678$} &
\multicolumn{2}{c}{$\phi_{\rm J}=0.697$}   \\
\hline
 (0,2,8,1)  &       2.8\%  &  (0,2,8,1)   &     5.1\%  &  (0,0,12)    &      14.2\%  \\
 (0,3,6,4)  &       2.5  &  (0,2,8,2)   &       4.4  &  (0,1,10,2)  &       7.0  \\
 (0,3,6,3)  &       2.4  &  (0,2,8)     &       4.1  &  (0,2,8)     &       6.3  \\
 (0,2,8,2)  &       2.3  &  (0,0,12)    &       3.9  &  (0,2,8,1)   &       5.7  \\
 (0,3,6,1)  &       2.0  &  (0,1,10,2)  &       3.7  &  (0,2,8,2)   &       4.6  \\
 (0,4,4,3)  &       1.9  &  (0,3,6,4)   &       3.2  &  (0,3,6)     &       3.7  \\
 (0,2,8)    &       1.8  &  (0,3,6,3)   &       2.8  &  (0,3,6,1)   &       3.0  \\
 (0,3,6,2)  &       1.7  &  (0,3,6,1)   &       2.8  &  (0,3,6,4)   &       2.4  \\
 (0,3,6)    &       1.6  &  (0,3,6)     &       2.3  &  (0,3,6,3)   &       2.1  \\
 (1,2,5,3)  &       1.4  &  (0,2,8,4)   &       2.1  &  (0,2,8,4)   &       2.0  \\
\hline
\hline
\end{tabular}
\end{center}
\caption{\label{table1} Most frequent Voronoi signatures and the corresponding average percentages in jammed configurations. The corresponding parent fluid packing fractions are $\phi_{\rm fluid}= 0.500 (N=8000)$, 0.618 ($N=8000$), and 0.655 ($N=1000$), respectively.}
\end{table}

Turning our attention to jammed packings, 
no distinguishable local order is detectable below $\phi_{\rm onset}$.
Figure \ref{fig:Q_and_LFS}(d) demonstrates that the sudden emergence of the icosahedral local order around $\phi_{\rm onset}$ in the equilibrium fluid is, again, inherited by the jammed packings.
It is thus tempting to attribute the increase of $\phi_{\rm J}$ along the J-line to the appearance of distorted icosahedral arrangements, which help maintaining an overall amorphous organization of the packings~\cite{klatt2014characterization}.
In Table~\ref{table1}, we report the occurrences of the most frequent Voronoi signatures in jammed packings at selected volume fractions.
We note that, on average, (0,0,12) cells have contact numbers $Z$ slightly smaller than 6 ($Z=5.4$).
We found that some less abundant signatures, such as (0,3,6) and (0,2,8), are associated to even smaller values of $Z$.
These trends suggest a correlation between $Z$ and the number of nearest neighbors, which we tentatively attribute to the size dispersity of the packings.

\begin{figure}
\begin{center}
\includegraphics[width=1.\columnwidth]{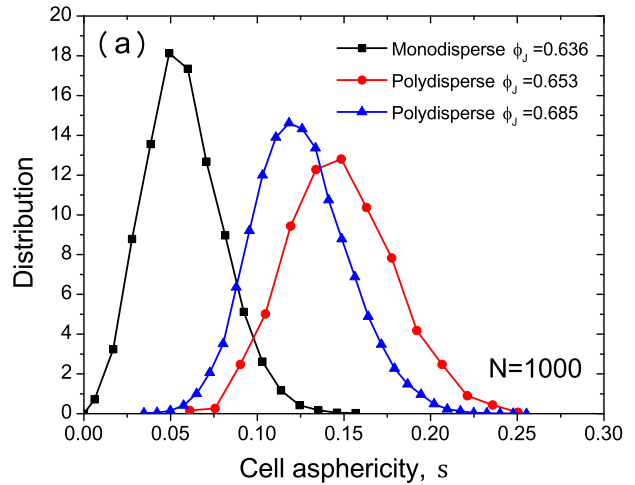}
\includegraphics[width=.95\columnwidth]{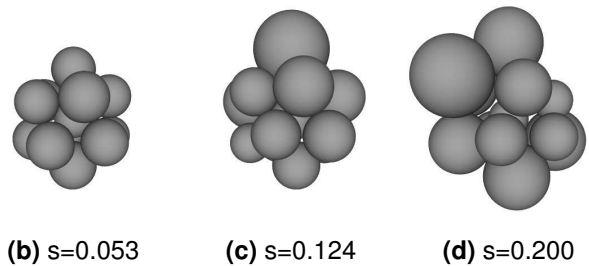}
\caption{(a): Distribution of asphericity parameter $s$ for 
polydisperse hard spheres 
for two different volume fractions, and for a monodisperse 
hard sphere system.
(b, c, d): Typical 
icosahedral structures with varying degree of asphericity $s$.}
\label{fig:asphericity}
\end{center}
\end{figure}  

The presence of icosahedral order and, more generally, polytetrahedral order 
in fluid and jammed hard sphere packings has been discussed before, see \textit{e.g.}~\cite{leocmach2012roles,Dunleavy_Wiesner_Yamamoto_Royall_2015,Anikeenko_Medvedev_2007,klatt2014characterization}.
The results presented above differ from earlier findings on two 
important aspects.
First, the amount of $(0,0,12)$ signatures found in our polydisperse packings substantially exceeds the one reported in monodisperse jammed packings~\cite{klatt2014characterization} and is also appreciably larger than the one reported in less polydisperse hard sphere at equilibrium~\cite{Dunleavy_Wiesner_Yamamoto_Royall_2015}.
However, because of the large size polydispersity, the local arrangements associated to $(0,0,12)$ signatures are often distorted and irregular.
Such issues are usually neglected in the analysis of icosahedral order in simple binary mixtures~\cite{coslovich2011locally}.

\begin{figure*}
\begin{center}
\includegraphics[width=1.\columnwidth]{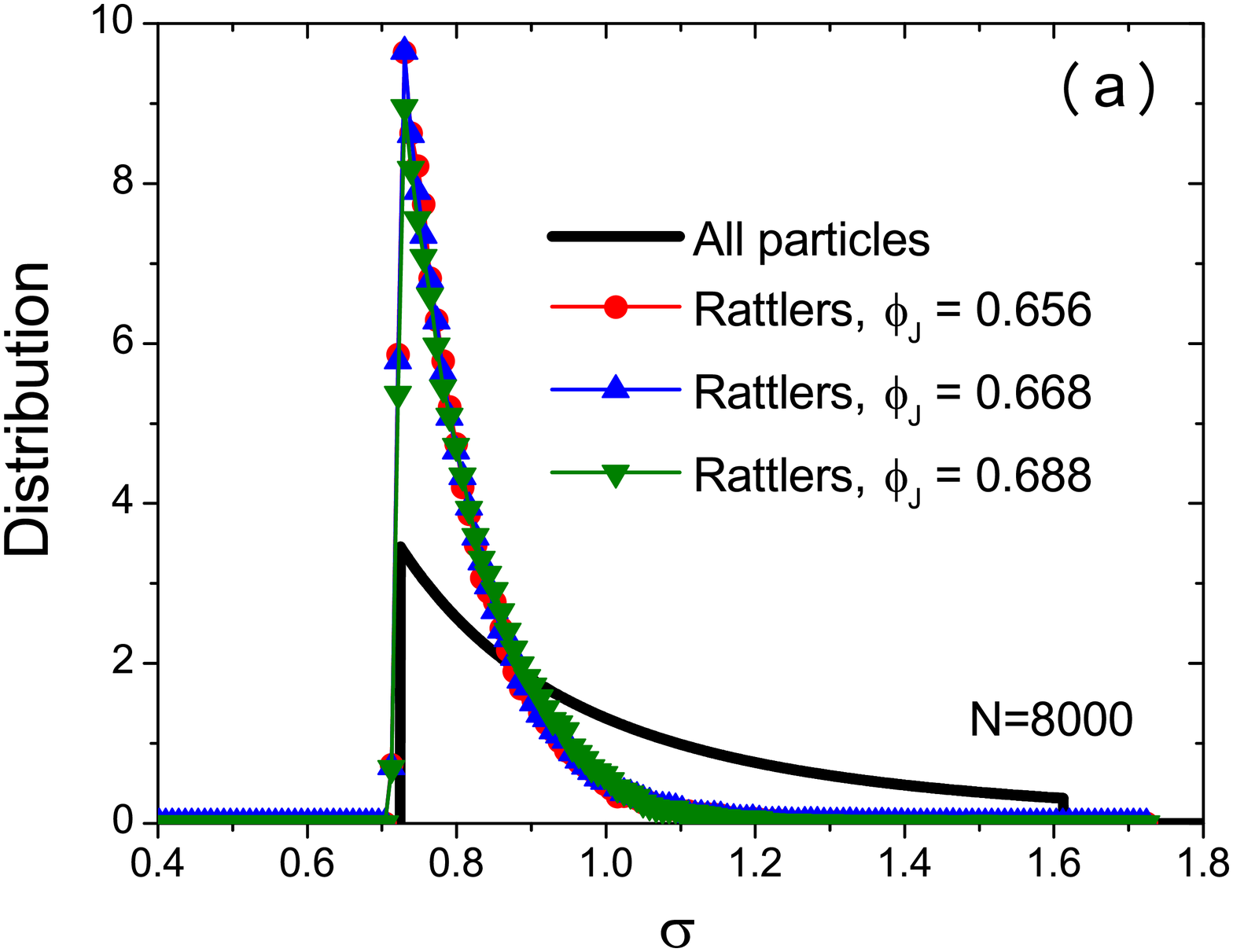}
\includegraphics[width=1.\columnwidth]{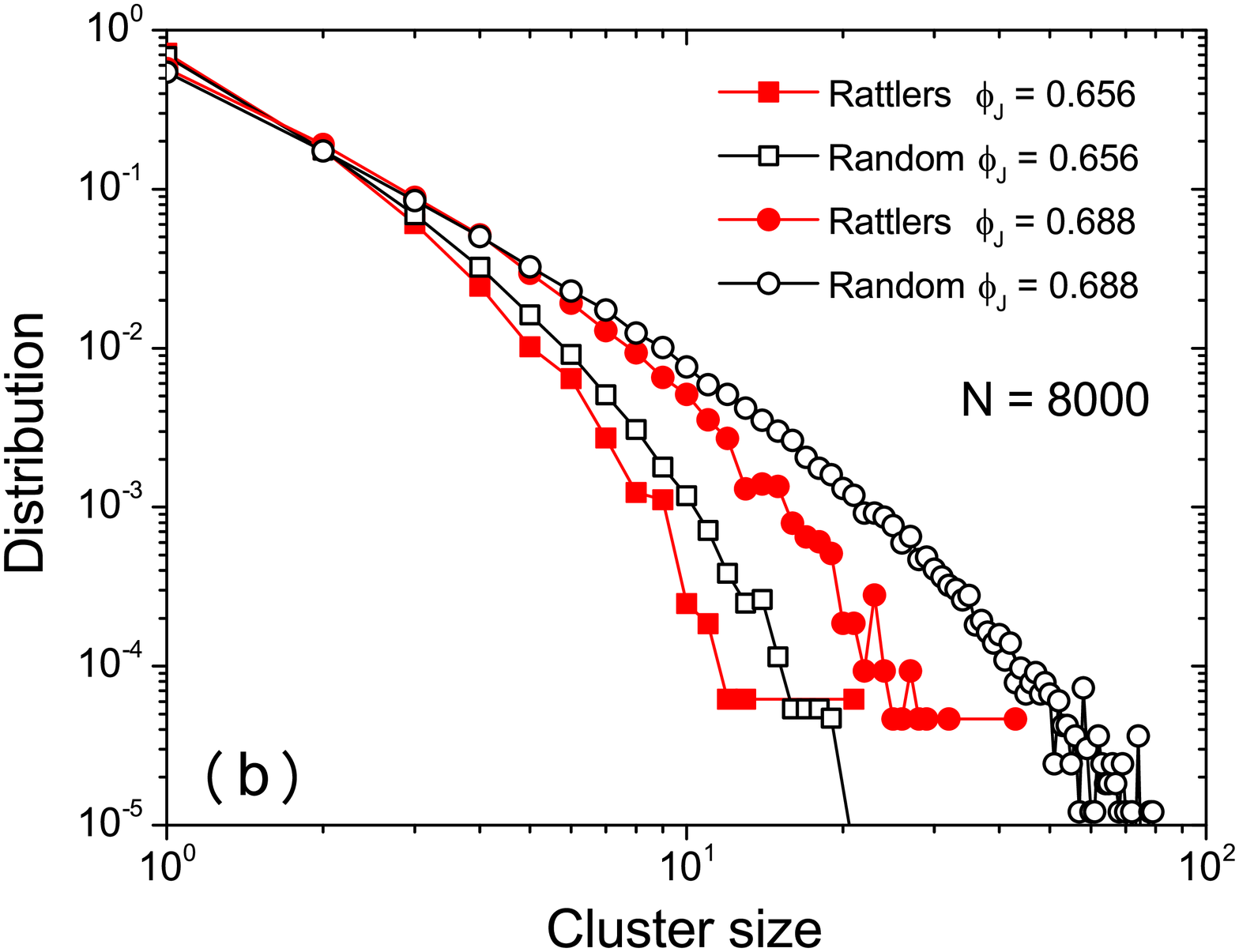}
\caption{
(a) Diameter distribution for all particles, $f(\sigma)$, and diameter distribution for rattlers only, $f_{\rm R}(\sigma)$.
(b) Cluster size distribution for the rattlers (closed symbols), compared
to the same fraction of randomly selected particles 
in the same packings (open symbols).}
\label{fig:rattler_analysis}
\end{center}
\end{figure*}  

To quantify the degree of asphericity of Voronoi cells, we analyzed the distribution of distances $r_{ij}=|{\bf r}_i-{\bf r}_j|$ separating a central particle $i$ to its neighbors $j$.
As a simple measure of cell asphericity, we compute the normalized standard deviation of the distances from particle~$i$, 
$$s_i = \frac{1}{\tilde{r}_i}\sqrt{\frac{1}{n_b(i)}\sum_{j=1}^{n_b(i)}(r_{ij} - \tilde{r}_i)^2} $$
where $\tilde{r}_i = (\sum_{j=1}^{n_b(i)} r_{ij}) /n_b(i)$ is the average nearest neighbors distance of particle $i$.
See Ref.~\cite{Schroder-Turk_Mickel_Schroter_Delaney_Saadatfar_Senden_Mecke_Aste_2010} for a more systematic approach based on Minkowski tensors. 
Of course $s\geq 0$ and the equality holds for a perfectly regular structure. 
In Fig.~\ref{fig:asphericity}(a) we show the 
probability distribution of the asphericity parameter, $P(s)$, 
of icosahedral structures for jammed packings for a low ($\phi_{\rm J}=0.653$) and a high ($\phi_{\rm J}=0.685$) packing fraction.
We find that the distribution is not very sensitive to the value of $\phi_{\rm J}$.
For comparison, we also include results for $P(s)$ measured in  
monodisperse hard sphere jammed packings. We obtained
the latter by compressing Poisson distributed configurations at $\phi_\textrm{fluid}=0.3$ using the algorithm described in Sec.~\ref{sec:compression}.
On average, icosahedral structures detected in polydisperse packings are more aspherical by about a factor two than those found in the monodisperse hard spheres.
We found that the asphericity of icosahedra in a glassy binary mixture with modest size ratio~\cite{wahnstrom}, for which icosahedral order is most pronounced~\cite{coslovich2007understanding}, is intermediate between those of the two systems in Fig.~\ref{fig:asphericity}(a).
Figures~\ref{fig:asphericity}(b-d) show representative icosahedral structures detected in our densest packings.
They are representative of different degrees of asphericity, ranging from fairly regular (b) to intermediate (c) and highly irregular (d) structures.

These results indicate that, despite the abundance of $(0,0,12)$ Voronoi cells, the nature of icosahedral order in polydisperse packings might be different from the one observed in simple binary mixtures.
In fact, icosahedral structures in highly polydisperse spheres are more distorted and inherently more irregular due to the increased compositional freedom.
In binary mixtures with sufficiently small size ratio icosahedral structures can strongly influence the local mobility of the particles in the supercooled regime~\cite{coslovich2007understanding,malins_identification_2013,hocky_correlation_2014}.
Further work is needed to assess the relevance of icosahedral local order in dense, highly polydisperse packings, in particular in the context of glass transition studies.

\subsection{Spatial organization of rattlers}

\label{sec:rattlers}

\begin{figure*}
\begin{center}
  \begin{tabular}{lr}
    \includegraphics[width=.8\columnwidth]{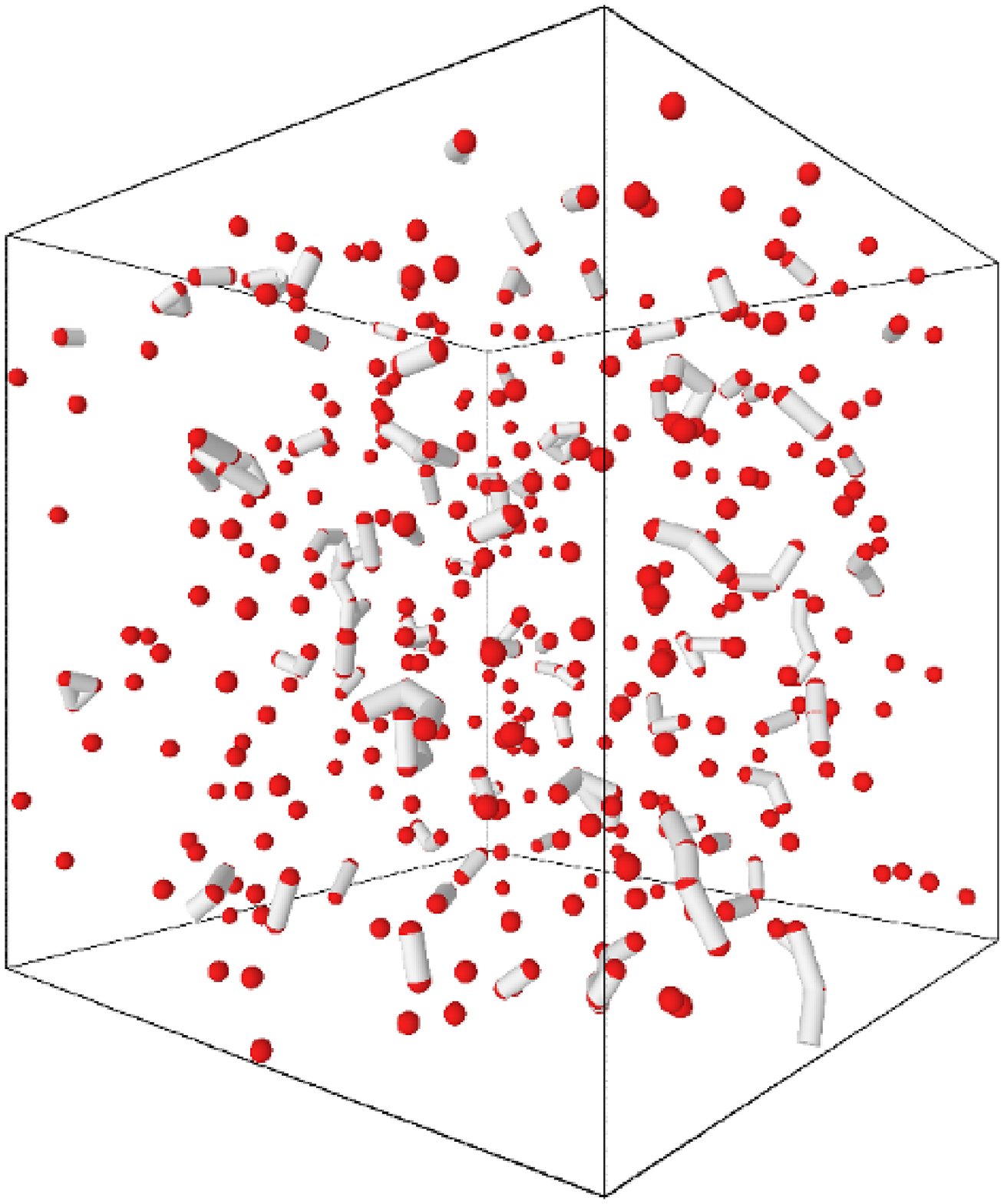} &
    \includegraphics[width=.8\columnwidth]{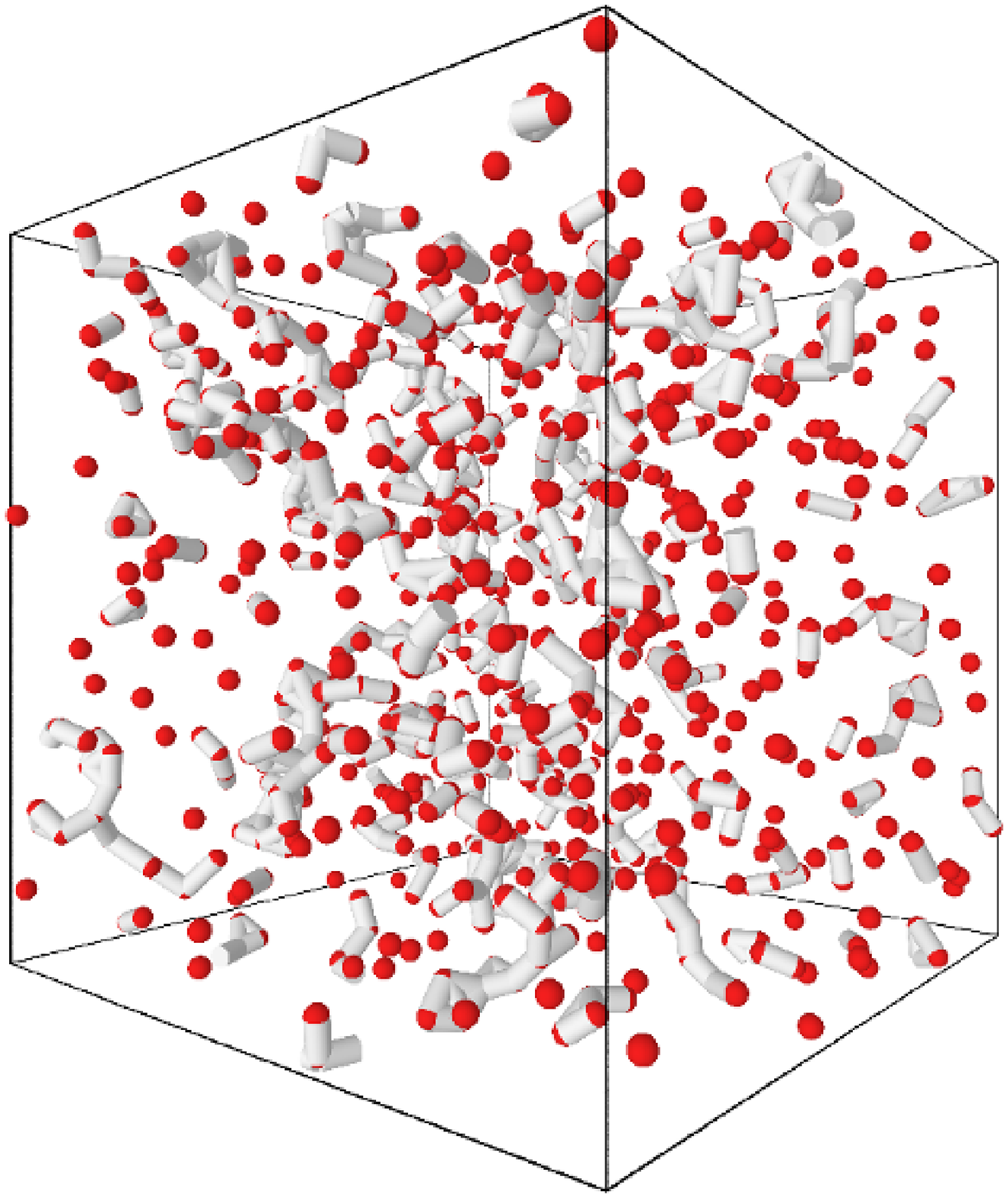}\\   
  \end{tabular}
\caption{
Typical snapshots of rattlers (red spheres) in jammed packings at $\phi_{\rm J}=0.656$ (left) and $\phi_{\rm J}=0.688$ (right). The white bonds connect neighboring rattlers that form a cluster.}
\label{fig:rattler_snapshots}
\end{center}
\end{figure*}

In Fig.~\ref{fig:contact_number}(b)
we have shown that the fraction of rattlers
grows steadily with increasing $\phij$. 
Naively, one might expect rattlers to occupy a larger free space, 
thus a large number of rattlers seems at odds with the increased efficiency of the packings.

First, we assess the size distribution of rattlers, $f_{\rm R}(\sigma)$, along the J-line, see Fig.~\ref{fig:rattler_analysis}(a).
By comparing $f_{\rm R}(\sigma)$ to the distribution of all particles, $f(\sigma)$, we find rattlers are mostly small particles, which is an intuitive result.
Interestingly, the shape of $f_{\rm R}(\sigma)$ hardly changes along the J-line, whereas the fraction of rattlers increases substantially.

Next, we inspect the spatial organization of rattlers.
To get a qualitative idea of their real space structure, we show in Figs.~\ref{fig:rattler_analysis}(c, d) some typical snapshots of rattlers found in jammed packings at $\phi_{\rm J}=0.656$ and $\phi_{\rm J}=0.688$.
Rattlers appear to be distributed in rather homogeneous way in this representation.
To get a more quantative picture, we follow previous previous work~\cite{atkinson2013detailed} and study the
spatial organization of rattlers into clusters. 
We define a rattler cluster as a group of rattlers where each rattler is neighbor to at least another member of the
cluster. As in the previous sections, neighbors are identified through a radical Voronoi tessellation.
Note that our analysis differs from the one of Ref.~\cite{atkinson2013detailed}, in which rattler clusters were classified according to the connectivity of their corresponding cages.

It is sometimes assumed that rattlers are randomly distributed in the
system~\cite{zachary2011hyperuniformPRL}.  To test this idea, we
randomly pick a fraction of particles with the same concentration as
the rattlers and compute the corresponding cluster size using the same
definition given above.  We show the distribution of rattler cluster
sizes for jammed states with a low ($\phi_{\rm J}=0.656$) and a high
($\phi_{\rm J}=0.688$) volume fraction in Fig.~\ref{fig:rattler_analysis}(b).
We find that the rattler clusters are smaller than the one formed
by the randomly chosen particles, for both low and high $\phi_{\rm J}$
values. This implies that overall the positions of the rattlers 
are only weakly correlated, displaying a slight tendency to form small
clusters~\cite{atkinson2013detailed}. This is confirmed by visual inspection of snapshots of jammed packings along the J-line, see Fig.~\ref{fig:rattler_snapshots} for representative examples. These relatively compact
clusters might lead to small volume fraction fluctuations and 
may also affect hyperuniform behavior, as discussed in Sec.~\ref{sec:conclusions}.

\section{Structure at the large scale}
\label{sec:global}

Here we consider the structure of the packings at the {large scale}.
The corresponding wave number regime is $k \to 0$, which means that
we quantify fluctuations at very large length scales.

\subsection{Hyperuniformity}

\begin{figure*}
\begin{center}
\includegraphics[width=0.95\columnwidth]{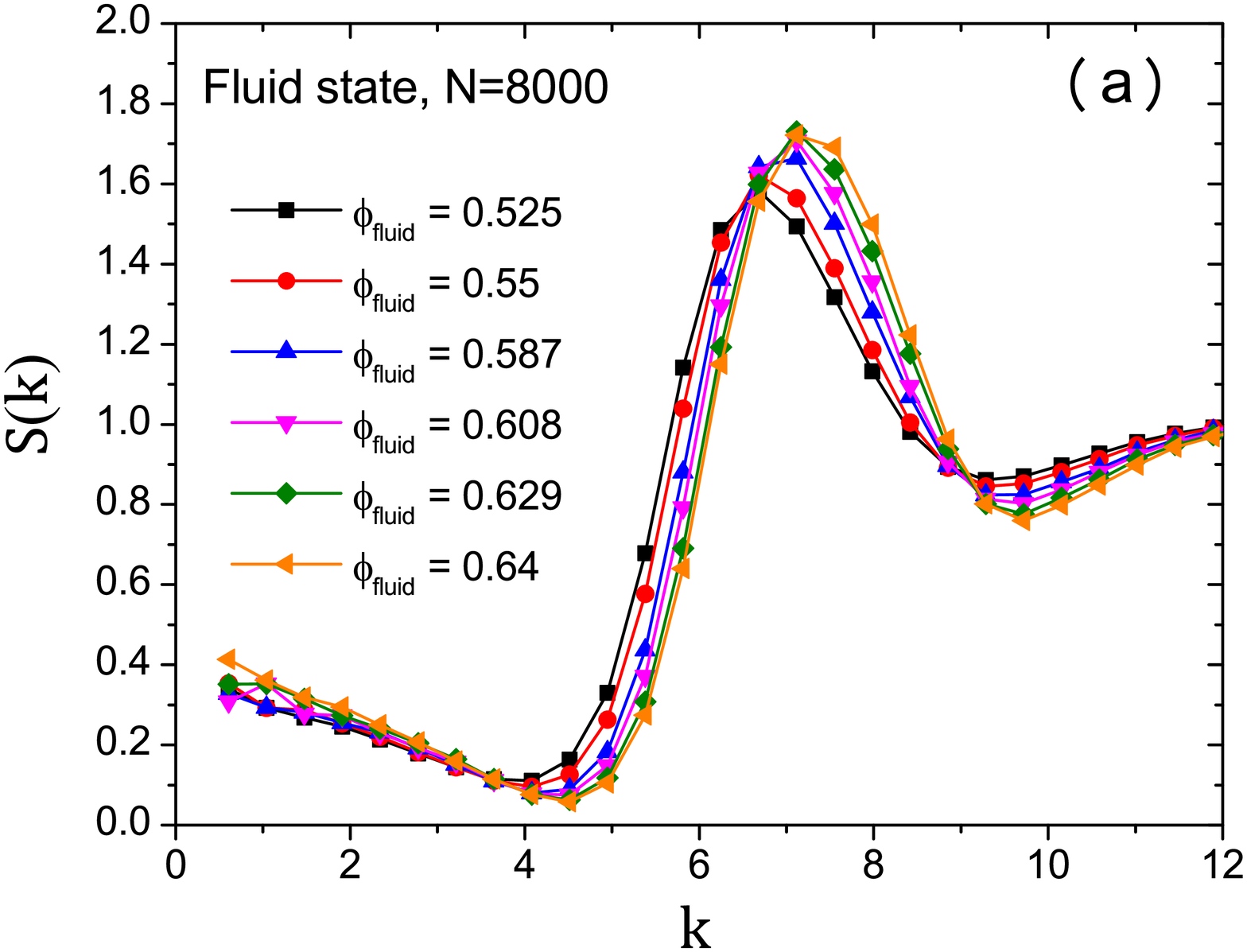}
\includegraphics[width=0.95\columnwidth]{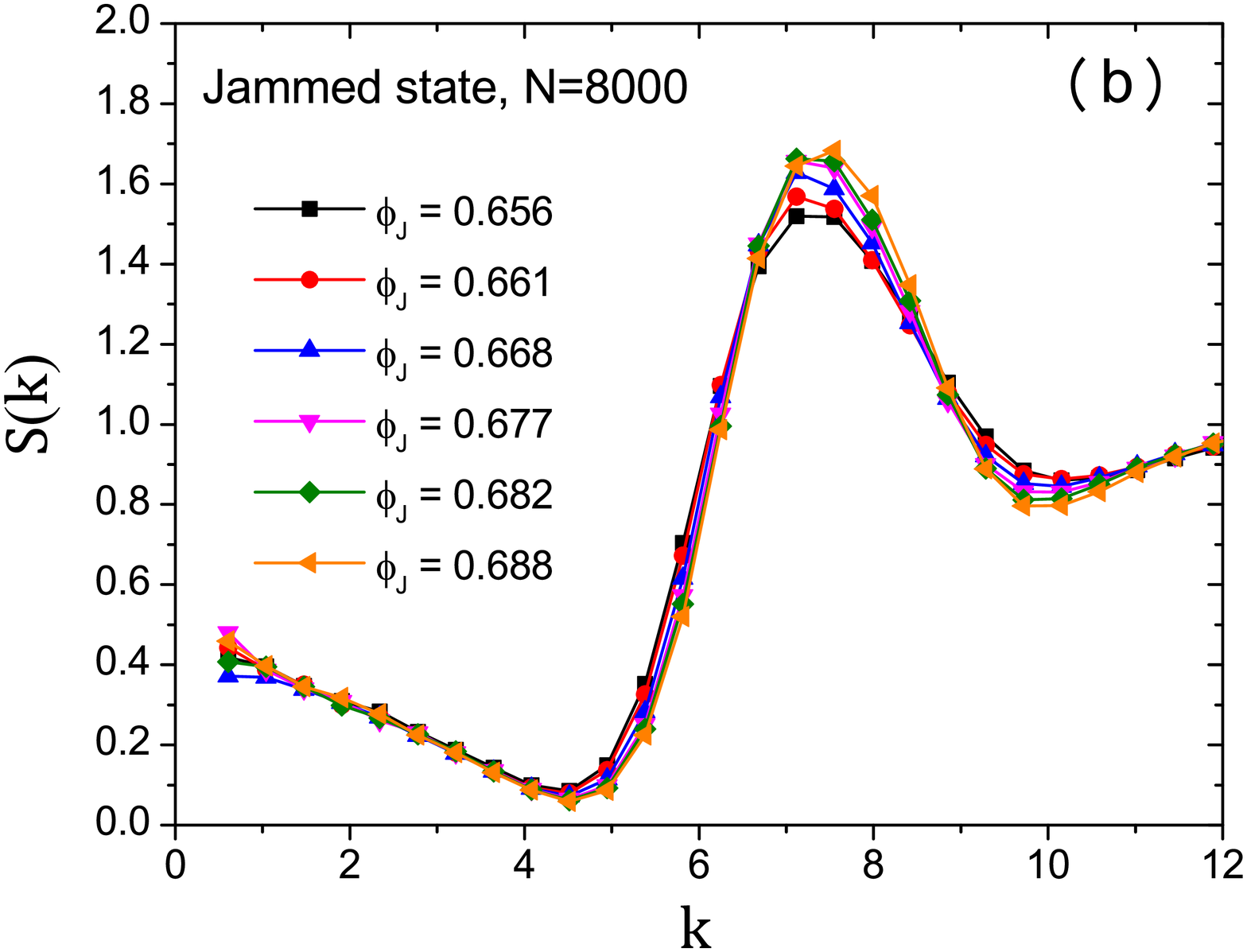}
\includegraphics[width=0.95\columnwidth]{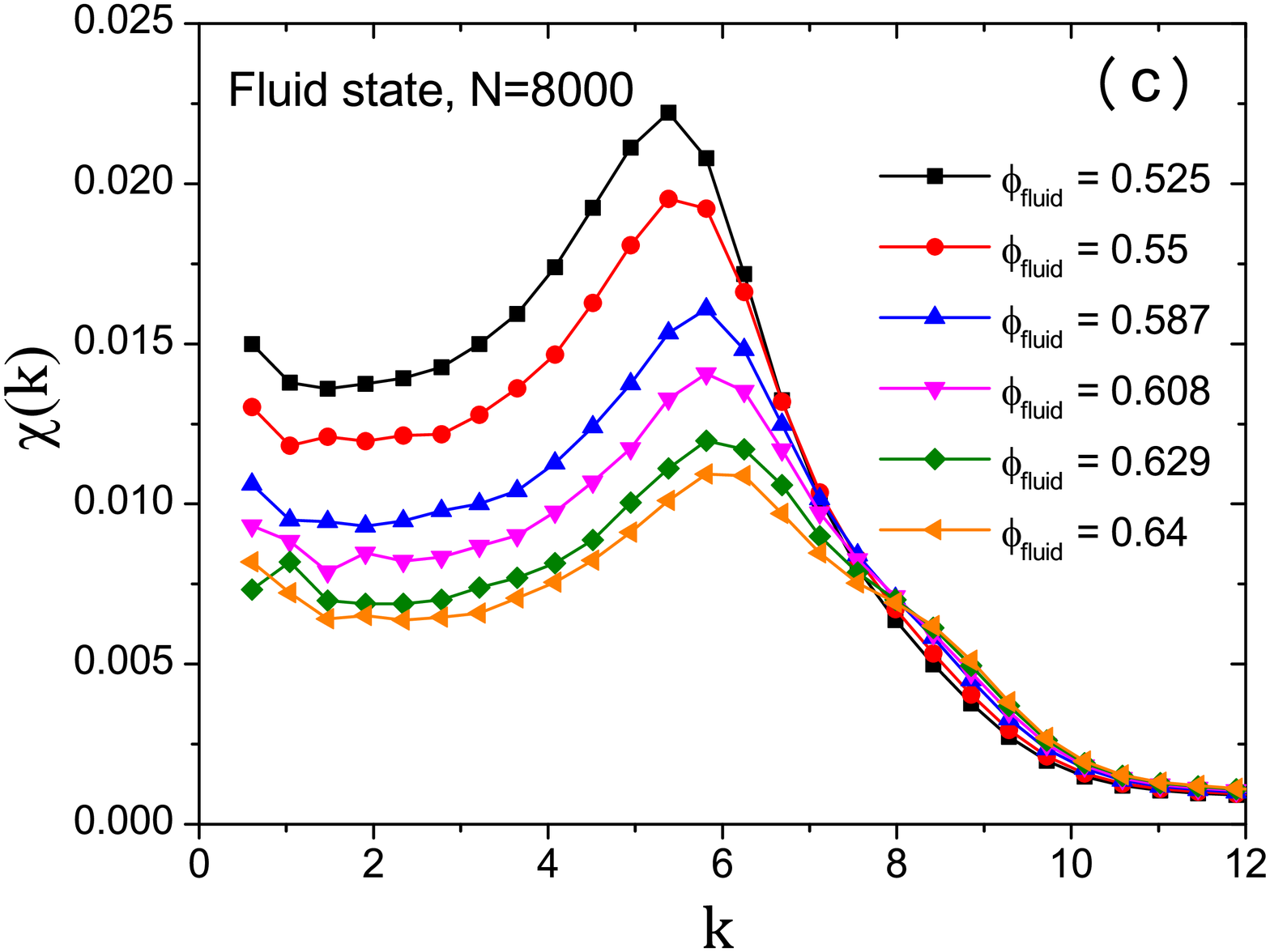}
\includegraphics[width=0.95\columnwidth]{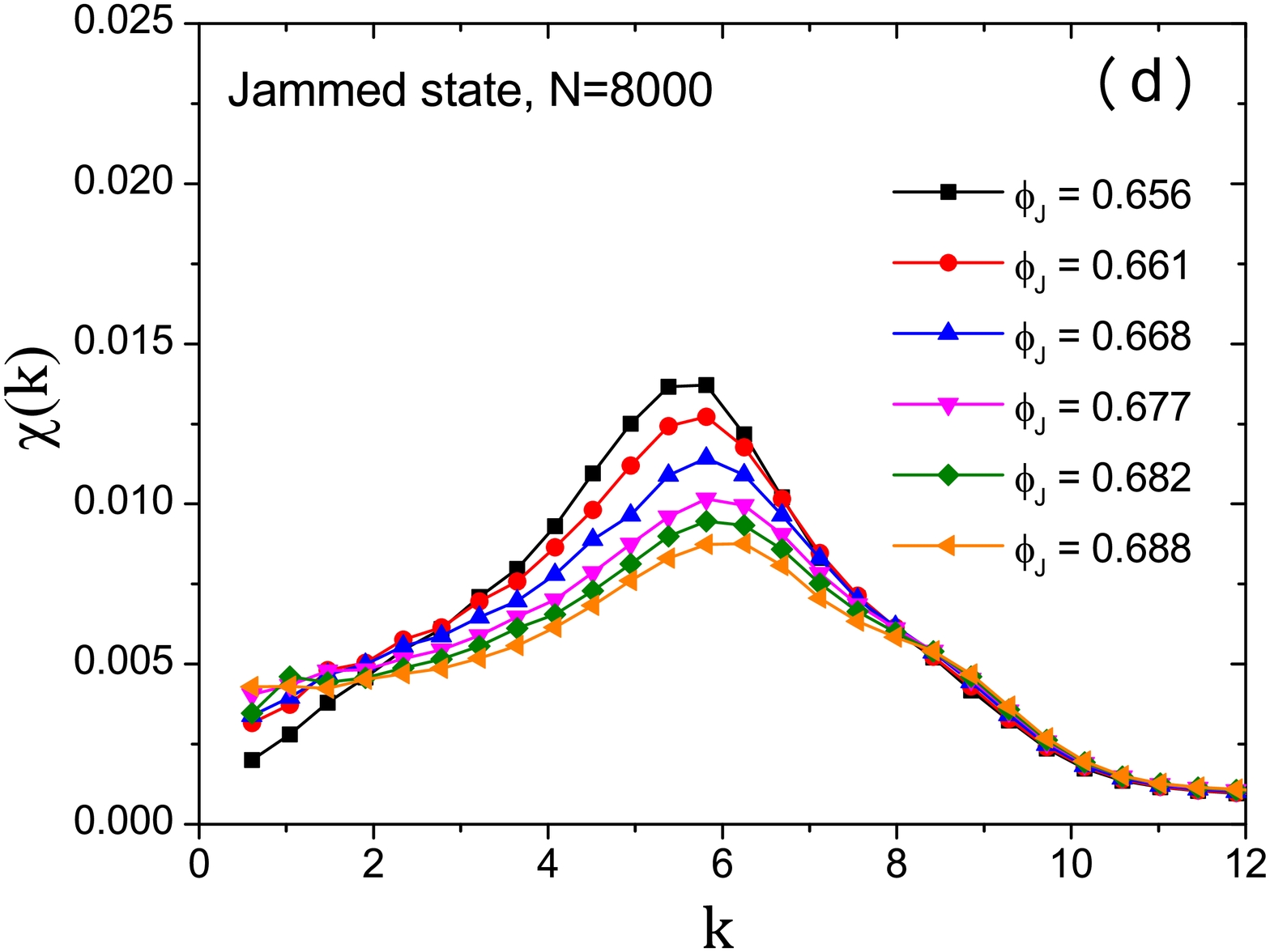}
\caption{
(a, b): Static structure factor $S(k)$ reflecting 
number density fluctuations for fluid (a) and jammed (b) states.
(c, d): Spectral density $\chi(k)$ reflecting 
volume fraction fluctuations for fluid (c) and jammed (d) states.
Nearly hyperuniform behavior expected 
for $\chi(k)$ is only observed for the lowest 
$\phij$ value in (d).}
\label{fig:Sofk_and_kai}
\end{center}
\end{figure*}  

Equilibrium fluids are characterized by a finite value of the static 
structure factor $S(k)$ at $k \to 0$, which is 
associated to a finite isothermal compressibility.
However, it is known that hard sphere packings at the jamming transition 
show ``unexpected'' (that is, more complex) density 
fluctuations~\cite{torquato2003local,donev2005unexpected}.
It was reported that 
$S(k)$ in monodisperse jammed packings obeys a surprising linear behavior
at low $k$, $S(k) \propto k$, which characterizes hyperuniform 
materials~\cite{torquato2003local,donev2005unexpected}.
It is sometimes assumed that hyperuniformity is related to 
the criticality of the jammed states~\cite{atkinson2016critical}, but the 
nature of this criticality and the connection to other signatures 
of jamming remain unclear~\cite{ikeda2015thermal,ikedaparisi}. 

When the system is a mixture of several components or 
is continuously polydisperse, the structure factor 
$S(k)$ of jammed states does not show a linearly vanishing behavior. 
Instead, $S(k)$ takes a finite value at $k \to 0$ just as in 
fluid states~\cite{xu2010effects,kurita2010experimental}.
We show $S(k)$ of our polydisperse system for both fluid and jammed 
states in Figs.~\ref{fig:Sofk_and_kai}(a, b).
These results confirm that the shape of $S(k)$ of jammed states is similar 
to the fluid ones, taking in particular a finite value, 
$S(k \to 0) \simeq 0.5$, for all jammed states along the J-line,
with no sign of a vanishing signal at low $k$. 
  
In Refs.~\cite{zachary2011hyperuniformPRL,zachary2011hyperuniformity}, 
the concept of hyperuniformity was generalized from density 
to volume fraction fluctuations to discuss the hyperuniformity 
of multi-components system and of particles with non-spherical shapes.
Thus, an appropriate observable to characterize the hyperuniformity of
polydisperse systems is the spectral density $\chi(k)$, which quantifies 
the volume fraction 
fluctuations~\cite{zachary2011hyperuniformPRL,zachary2011hyperuniformity}. 
It is defined as 
\begin{equation}
\chi({\bf k}) = \frac{1}{V} \left\langle I_{{\bf k}} I_{-{\bf k}} \right\rangle,
\end{equation}
where $I_{\bf k}$ is 
the Fourier transform of the indicator function $I({\bf r})$.
For spherical particle systems, $I({\bf r})$ is given by
\begin{equation}
I({\bf r})=\sum_{i=1}^N \theta(R_i - |{\bf r}-{\bf r}_i|),
\end{equation}
where $R_i$ is the radius of the $i$-th particle, $R_i=\sigma_i/2$.
For homogeneous isotropic systems in $d=3$, $I_{\bf k}$ becomes  
\begin{equation}
I_{\bf k} = \sum_{i=1}^N \frac{4 \pi}{k^3} \left[ \sin(k R_i) - (kR_i) \cos(kR_i) \right] e^{-i {\bf k} \cdot {\bf r}_i}.
\end{equation}
Note that alternative definitions of the volume fraction 
fluctuations in the reciprocal space can be applied to study 
hyperuniformity in jammed packings~\cite{berthier2011suppressed,wu2015search}.

In Figs.~\ref{fig:Sofk_and_kai}(c, d), we show our numerical results 
for $\chi(k)$ for both fluid and jammed states.
Before discussing $\chi(k)$ of jammed states at small $k$, 
where signs of hyperuniformity may be found, it is instructive to 
study $\chi(k)$ for the fluid states and its volume fraction dependence.
We find that $\chi(k)$ for fluid states has a broad peak near 
$k \sim 2 \pi/ \overline{\sigma}$ and a flat plateau at smaller 
$k$~\cite{klatt2016characterization}. In addition, the overall 
amplitude of  
$\chi(k)$ decreases as the volume fraction $\phi_{\rm fluid}$ increases 
at $k \lesssim 2 \pi/ \overline{\sigma}$, reflecting that 
volume fraction fluctuation are suppressed in order to achieve 
denser particle packings in equilibrium.
This observation contrasts with the well-known increase of the 
first peak of $S(k)$ with increasing the density 
in dense fluids~\cite{binder2011glassy} which instead reflects the 
slight increase in local scale structural correlations
as the density is increased. 

In jammed states, the peak height of $\chi(k)$ is 
much smaller than in corresponding fluid states, and the 
peak amplitude decreases as 
$\phi_{\rm J}$ increases, reflecting again that further optimization 
of local configurations has been realized.
Note that $\chi(k)$ displays a rather strong 
volume fraction dependence in both 
fluid and jammed states, which contrasts with the relatively 
weak density dependence of $S(k)$ over the same range. 
This insensitivity of $S(k)$ in supercooled liquids 
is traditionally taken as a hallmark
of glass physics~\cite{binder2011glassy}. 
However, our results demonstrate that $\chi(k)$ is a more sensitive probe of
structural changes at the level of two-point correlations. 
This suggests that $\chi(k)$ might be a good observable 
to characterize the equilibrium structure of dense hard sphere systems.

\begin{figure}
\begin{center}
\includegraphics[width=1.0\columnwidth]{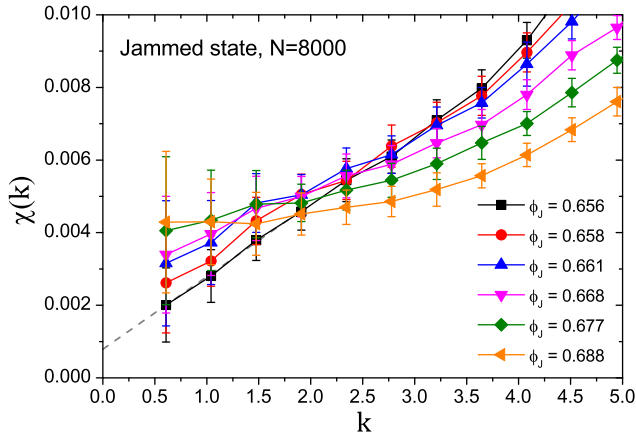}
\caption{Zoom in the small $k$ region of the spectral 
density $\chi(k)$ for jammed states shown in Fig.~\ref{fig:Sofk_and_kai}(d). 
The error bars represent the standard deviation obtained 
from sample to sample fluctuations.
The gray dashed line is a linear fit for $\phi_{\rm J}=0.656$, 
which extrapolates to $\chi(k \to 0) = 7.9 \times 10^{-4}$ for 
the lowest $\phij$. Instead $\chi(k \to 0) = 4.2 \times 10^{-3}$ for the 
largest $\phij$.}
\label{fig:kai_zoom}
\end{center}
\end{figure}  

Having described the main features of $\chi(k)$, we now 
concentrate on the low-$k$ behavior in jammed states.
A zoom in this region is shown in Fig.~\ref{fig:kai_zoom}, to
investigate more carefully the possible existence of a
hyperuniform behavior in this regime.
At the lowest $\phi_{\rm J}$ along the J-line for $N=8000$, 
$\phi_{\rm J}=0.656$, $\chi(k)$ linearly decreases with decreasing
$k$, 
as observed in previous numerical works~\cite{zachary2011hyperuniformPRL,zachary2011hyperuniformity,wu2015search,atkinson2016critical}. 
By fitting this linear behavior (shown with the dashed line), 
we can extrapolate the limit $k \to 0$ and we obtain a small 
finite value, $\chi(k \to 0) = 7.9 \times 10^{-4}$ for this volume fraction.
Therefore, one might conclude that the system at $\phi_{\rm J}=0.656$ 
is very close to being hyperuniform~\cite{atkinson2016critical}.
However, as $\phi_{\rm J}$ is increased from its lowest value, $\chi(k)$ 
quickly deviates from this linear behavior, and it even becomes flat, 
within our error bars, at low $k$. 
Simultaneously, $\chi(k \to 0)$ obtained by linear extrapolation increases 
systematically from $\chi(k \to 0) \approx 7.9 \times 10^{-4}$
to $\chi(k\to0) \approx 4.2 \times 10^{-3}$
with increasing $\phi_{J}$, suggesting 
that deviations from hyperuniformity become stronger 
when $\phi_{\rm J}$ increases.

Recently, deviations from hyperuniformity were 
reported 
in both two~\cite{wu2015search} and three~\cite{ikeda2015thermal,ikedaparisi} 
spatial dimensions.
Both works employed rapid quenches from fully random configurations using 
energy minimization protocols. 
These studies showed that $\chi(k)$ (and thus $S(k)$ in monodisperse systems) 
linearly decreases with decreasing $k$, but it always saturates at a 
certain $k^*$ ($k^* \sim 0.2$ in Ref.~\cite{wu2015search} and 
$k^* \sim 0.4$ in Refs.~\cite{ikeda2015thermal,ikedaparisi}), 
indicating that 
hyperuniformity is eventually avoided at sufficiently small $k$.
Ref.~\cite{atkinson2016critical} attributes the breakdown of 
hyperuniformity at very small $k$ to a lack of numerical accuracy in conventional jamming 
algorithms.
It is argued that producing truly jammed, hyperuniform states is a highly 
difficult numerical task and thus the 
observed saturation at $k^*$ using energy minimization 
could stem from numerical inaccuracy of a given algorithm.

However, the systematic deviation of $\chi(k)$ at denser $\phi_{\rm J}$ of 
our system shown in Fig.~\ref{fig:kai_zoom} presumably has a different
origin, because the system size we use, $N=8000$, 
is not large enough to detect the saturation at $k^*$ reported in 
previous work. 
Also, the systematic increase of $\chi(k \to 0)$ is observed for a given
system size and within a given numerical algorithm.
Therefore, the strong deviations from hyperuniformity reported here
must have a physical, rather than a computational, origin. 

Our jammed packings at various $\phij$ are obtained with a similar degree 
of precision, in particular isostaticity is equally well satisfied 
for all configurations along the J-line. However, 
they display an increasing tendency to depart 
from hyperuniformity as $\phij$ is increased. Therefore, we
conclude that the jamming criticality, which is well-obeyed all
along the J-line, and hyperuniformity, which is increasingly
violated, are distinct concepts. 
This conclusion is consistent with the fact that the linear dependence 
of $\chi(k)$ (and $S(k)$), which might be interpreted as 
a ``precursor'' of hyperuniformity~\cite{hopkins2012nonequilibrium}, 
is robustly observed at densities well below~\cite{hopkins2012nonequilibrium} and above~\cite{ikeda2015thermal,wu2015search} 
$\phi_{\rm J}$ and even in the presence of thermal fluctuations~\cite{ikeda2015thermal}, 
whereas the jamming criticality originating from isostaticity 
is very quickly erased in similar conditions~\cite{ikeda2013dynamic}.

Furthermore, the nearly hyperuniform behavior observed at the lowest 
$\phi_{\rm J}$ in this work implies that hyperuniformity, if it were to
be observed, should characterize the lowest end of the 
J-line~\cite{ciamarra2010recent,kumar2016memory}.
Although the ``maximally random jammed
state''~\cite{torquato2010jammed} and the lowest end point of the J-line 
are conceptually distinct, our observations lead us to speculate 
that these two concepts may be identical~\cite{parisi2010mean},
and may both display the strongest (although 
presumably still imperfect) signature 
of hyperuniformity. This conclusion, based on our numerical findings,
also directly contradicts an opposite theoretical prediction recently made in 
Ref.~\cite{coniglio2017universal}.

A hypothesis that could explain the present observations 
is that the increasing deviation from hyperuniformity observed
on increasing $\phij$ correlates with the increasing number of rattlers
in the corresponding packings. Rattlers can be seen as ``defects''
in the volume fraction field, and a finite fraction of rattlers could induce
a finite amount of volume fraction fluctuations at large scale. This 
hypothesis is hard to test directly, as we cannot independently
vary the fraction of rattlers and the degree of hyperuniformity 
in hard sphere packings.
We notice that the fraction of rattlers increases by only a factor of 2 while the limit $\chi(k\to0)$ increases by about an order of magnitude, which suggests that rattlers may not be the central explanation for this observation.
This conclusion was indirectly tested in Ref.~\cite{wu2015search}, which 
showed that upon compression of jammed packings the number of rattlers
decreased but the behavior of $\chi(k)$ was essentially unchanged.
We also confirmed this behavior in our simulations (not shown).
We will discuss this issue further in Section~\ref{sec:conclusions}. 

\subsection{Finite-size fluctuations of the critical 
density of jamming}

\begin{figure}
\begin{center}
\includegraphics[width=1.0\columnwidth]{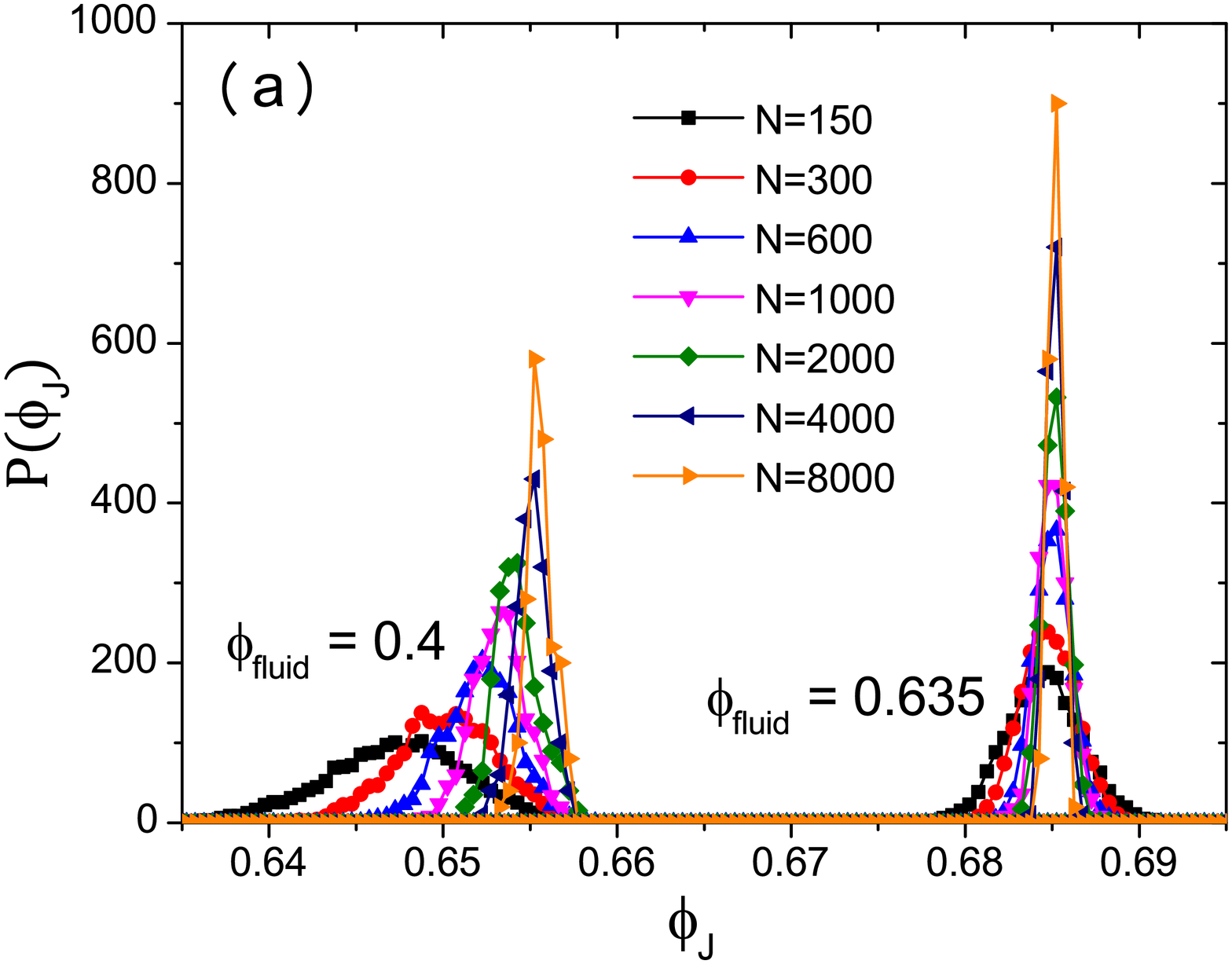}
\includegraphics[width=1.0\columnwidth]{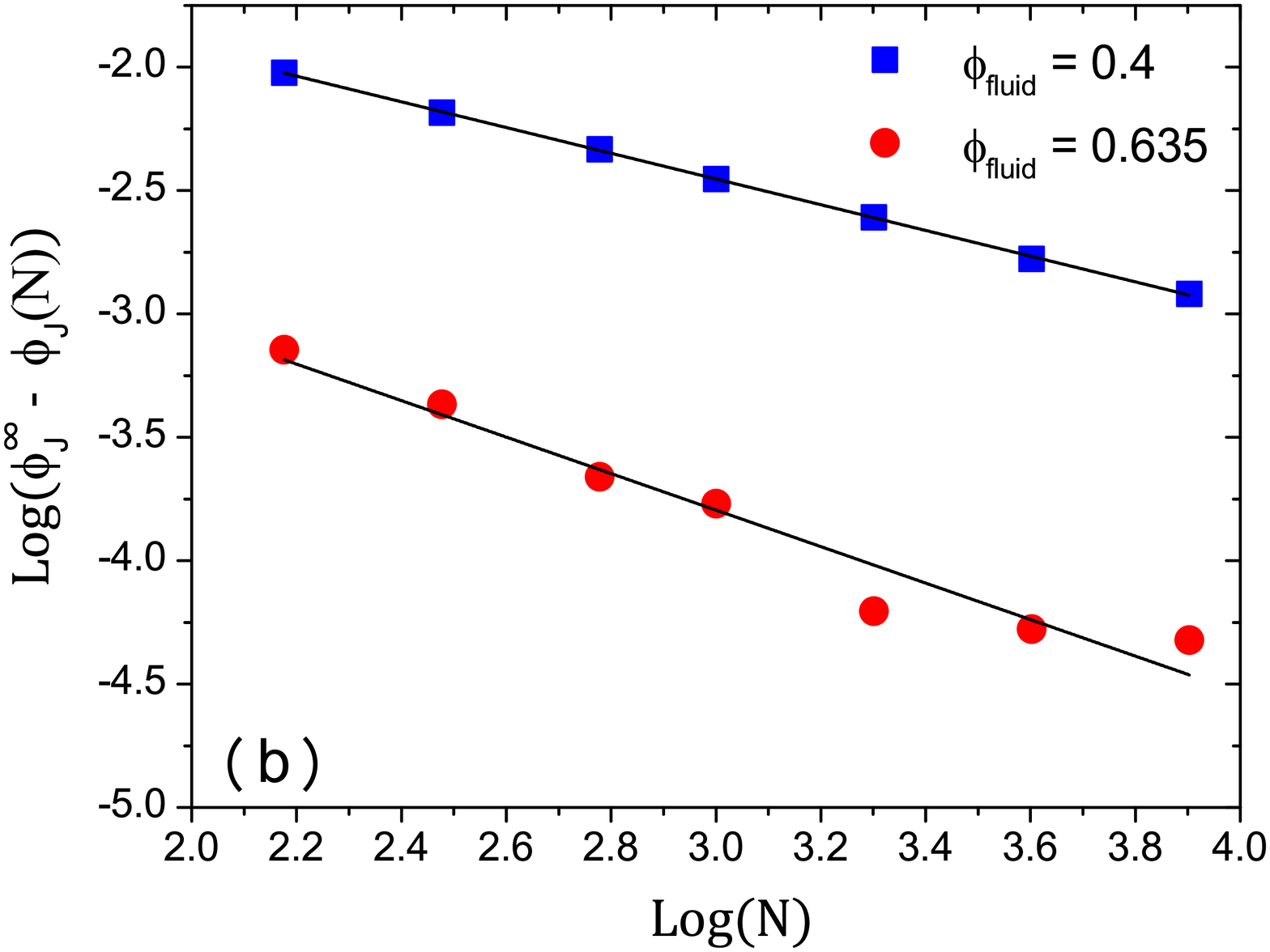}
\caption{
(a): Probability distribution function of $\phi_{\rm J}$ for 
several $N$ obtained by compressing dilute ($\phi_{\rm fluid}=0.4$) and
dense ($\phi_{\rm fluid}=0.635$) equilibrium fluid configurations.
(b): Finite-size scaling plot according to Eq.~(\ref{eq:phiJ_scaling}).
Finite-size effects on the quantity $\phij(N)$
are considerably suppressed for $\phi_{\rm fluid} = 0.635$.} 
\label{fig:finite_size_scaling}
\end{center}
\end{figure}  

Second order phase transitions are 
usually associated with diverging length scales~\cite{nishimori2010elements}. 
Finite-size scaling represents a powerful tool to understand the 
nature of such transitions and of the corresponding diverging length scales.
Although several distinct important length scales have been identified 
for the jamming transition~\cite{silbert2005vibrations,drocco2005multiscaling,reichhardt2014aspects,goodrich2013stability,schoenholz2013stability,ellenbroek2009jammed,olsson2007critical,hopkins2012nonequilibrium,ikeda2013dynamic,ikeda2015one,ikeda2015thermal}, the particular length scale responsible for the finite-size 
effects of $\phi_{\rm J}$ is not well 
understood~\cite{o2003jamming,drocco2005multiscaling,vaagberg2011finite}.

A qualitatively different aspect of the jamming transition compared to usual critical phenomena is the fact that the transition point is not 
unique and is protocol-dependent~\cite{goodrich2016scaling}.
Thus, we examine the finite-size effect of $\phi_{\rm J}$ for 
two different protocols which produce different averaged $\phi_{\rm J}$ values.
In practice, we compare results of two protocols, the non-equilibrium 
compression from dilute ($\phi_{\rm fluid}=0.4$) and dense 
($\phi_{\rm fluid}=0.635$) fluids, which produce low ($\phij \simeq 0.657$)
and high ($\phij \simeq 0.685$) jamming volume fractions. We expect 
a smooth evolution of the behavior with $\phij$, but these
measurements are numerically demanding, so we limit ourselves
to only two different critical points on the extremes of the J-line.

We show the probability distribution function $P(\phi_{\rm J})$ for several system sizes $N$ in Fig.~\ref{fig:finite_size_scaling}(a).
For the compression of the dilute fluid, $\phi_{\rm fluid}=0.4$, a broad Gaussian shape of $P(\phi_{\rm J})$ is obtained for the smallest system size. 
As $N$ increases, the distribution becomes narrower and the position of the
peak shifts to higher volume fractions, as reported 
previously~\cite{o2003jamming,xu2005random,gao2006frequency}.  

For the compression of the dense fluid, $\phi_{\rm fluid}=0.635$, the width of the distribution for a given $N$ is slightly narrower than the one for $\phi_{\rm fluid}=0.4$~\cite{vaagberg2011glassiness,rieser2016divergence}, but the 
distribution again becomes more sharply peaked as $N$ increases.
Remarkably, the peak position hardly shifts with $N$, in 
strong contrast with $\phi_{\rm fluid}=0.4$.
Therefore, for $\phi_{\rm fluid}=0.635$, the finite size effect in terms of 
the mean value of $\phi_{\rm J}$ is significantly suppressed
and $\phi_{\rm J}(N)$ quickly approaches the thermodynamic 
value $\phi_{\rm J}^{\infty}$. For both protocols, the latter value can be 
extracted using standard finite-size scaling, 
\begin{equation}
\phi_{\rm J}(N) = \phi_{\rm J}^{\infty}  - \delta N^{-1/\nu d},
\label{eq:phiJ_scaling}
\end{equation}
where $\delta$ and $\nu$ are fitting parameters, 
and $d$ is the number of spatial dimensions.
Note that we use the mean value for $\phi_{\rm J}(N)$ instead of the peak value~\cite{o2003jamming}, since we do not have enough statistics to determine the peak position with high accuracy.
The results of this fit are shown in Fig.~\ref{fig:finite_size_scaling}(b).
We obtain $\phi_{\rm J}^{\infty}=0.65698 \pm 0.00009$ and $\nu=0.64 \pm 0.01$ for $\phi_{\rm fluid}=0.4$, and $\phi_{\rm J}^{\infty}=0.68548 \pm 0.00002$ and $\nu=0.41 \pm 0.04$ for $\phi_{\rm fluid}=0.635$, respectively.
The obtained $\nu$'s are slightly different from previous reports~\cite{o2003jamming,ciamarra2010disordered}.
Note that $\nu$ for $\phi_{\rm fluid}=0.4$ is compatible with very recent numerical work~\cite{coniglio2017universal}. 
However, we do not wish to discuss these values quantitatively because $\nu$ might be protocol or algorithm dependent, and corrections to scaling should 
be included before drawing any strong conclusions~\cite{vaagberg2011finite}.
We do not treat these corrections  to scaling 
because they require huge statistics.
Our main result here is more qualitative; 
finite size effects on $\phi_{\rm J}$ are significantly suppressed 
for the compression from the dense fluid. 
The prefactor $\delta$ in Eq.~(\ref{eq:phiJ_scaling}) is more than
10 times smaller for the compression of the dense fluid, as 
can be directly seen in the data shown 
in Fig.~\ref{fig:finite_size_scaling}(b). The smallness 
of the finite-size effect in fact makes a quantitative 
determination of the critical exponent very difficult
for the largest $\phij$ values~\cite{ozawa2012jamming}.
Our data conclusively demonstrate that the J-line is not due to a finite size effect and that it remains instead well-defined in the thermodynamic limit.

\begin{figure}
\begin{center}
\includegraphics[width=1.\columnwidth]{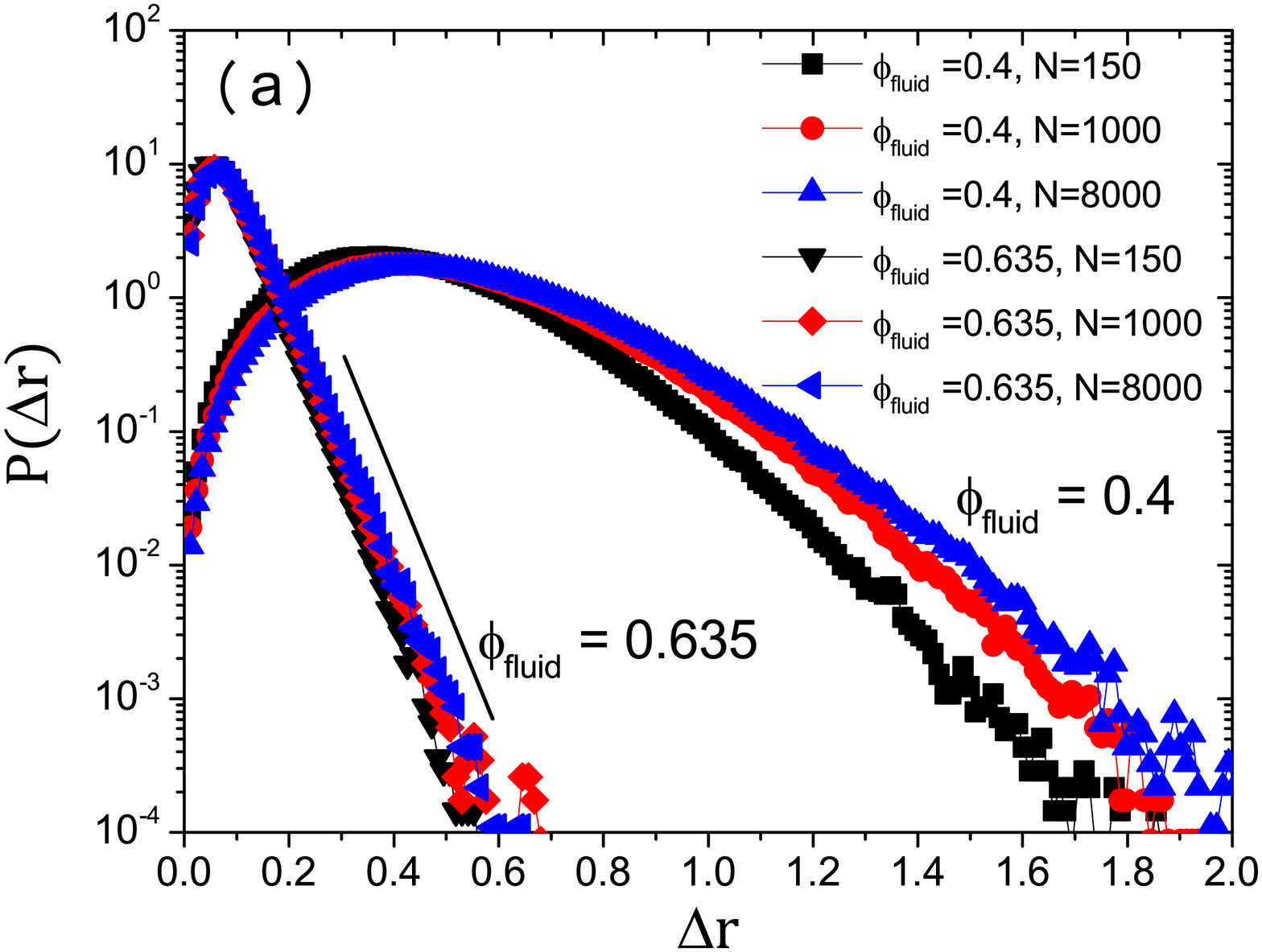}
\includegraphics[width=1.\columnwidth]{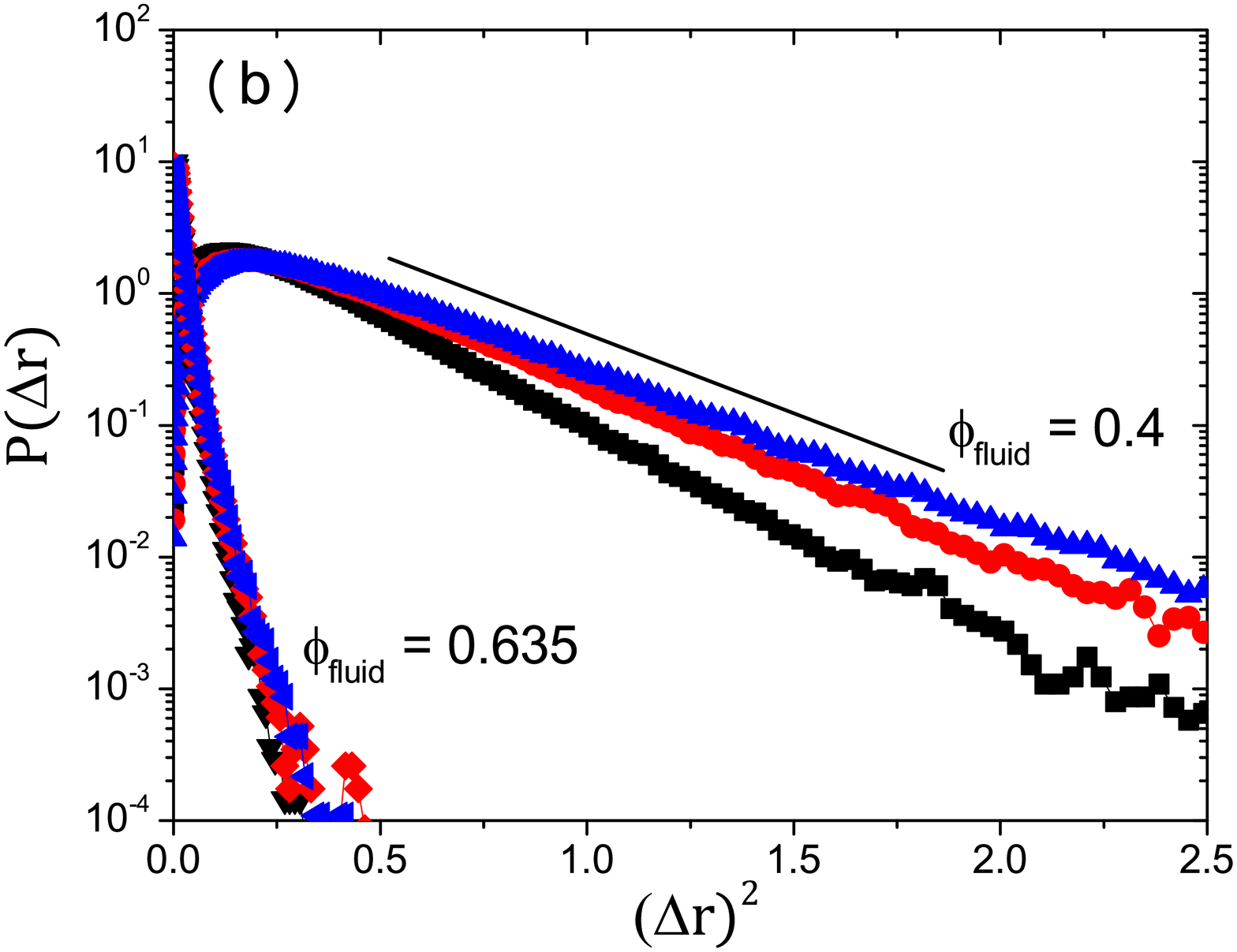}
\caption{
(a): Probability distribution function $P(\mathit{\Delta} r)$ of 
particle displacements during a compression, 
$\mathit{\Delta} r = | {\bf r}_i^{\rm jam} - {\bf r}_i^{\rm fluid}|$. 
The straight line corresponds to $P(\mathit{\Delta} r) \propto \exp[ - a \mathit{\Delta} r]$, where $a$ is a constant.
(b): The same distributions $P(\mathit{\Delta} r)$ 
as a function of the quantity $(\mathit{\Delta} r)^2$.
The straight line corresponds to $P(\mathit{\Delta} r) \propto \exp[ - b (\mathit{\Delta} r)^2]$, where $b$ is a constant.
}
\label{fig:P(Dr)_vs_Dr}
\end{center}
\end{figure}  

To get more physical insights into the finite-size effects for $\phi_{\rm J}$, we monitor the probability distribution function $P(\mathit{\Delta} r)$ of 
single particle displacements during the compression from the 
fluid to the jammed states, $\mathit{\Delta} r = |  {\bf r}_i^{\rm jam} - {\bf r}_i^{\rm fluid}|$, where ${\bf r}_i^{\rm fluid}$ and ${\bf r}_i^{\rm jam}$ are the positions of 
particle $i$ in the equilibrium parent fluid configuration and in the 
corresponding jammed configuration, respectively. 
The distribution $P(\mathit{\Delta} r)$ is computed for all the particles (including rattlers).
$P(\mathit{\Delta} r)$ quantifies how much the particles need to 
move or rearrange during the compression until the system is jammed.
In Figs.~\ref{fig:P(Dr)_vs_Dr}(a, b), 
we show $P(\mathit{\Delta} r)$ as a function 
of either $\mathit{\Delta} r$ or $(\mathit{\Delta} r)^2$, 
for three different system sizes, $N=150$, $1000$, and $8000$.
For $\phi_{\rm fluid}=0.4$, $P(\mathit{\Delta} r)$ has a rather long tail, indicating that the particles may perform large displacements 
before finding the final jammed configuration.
This tail can be fitted by a Gaussian form, $P(\mathit{\Delta} r) \propto \exp[ - b (\mathit{\Delta} r)^2]$, where $b$ is a constant 
[see Fig~\ref{fig:P(Dr)_vs_Dr}(b)], which suggests that there is no 
strong positional correlation between before and after compression.
Importantly, a noticeable finite-size effect is found in these tails,
in the sense that smaller systems are characterized by 
smaller particle displacements.
This observation provides a possible explanation for the observed 
finite-size effect in $\phi_{\rm J}$: particles in smaller systems
do not explore space as they do in large systems and thus get jammed
in less optimized ({\it i.e.} less dense) configurations, leading to smaller 
$\phij$, as observed in Fig.~\ref{fig:finite_size_scaling}. 

Interestingly, for $\phi_{\rm fluid}=0.635$, the tail 
in $P(\Delta r)$ is significantly
suppressed compared to the one for $\phi_{\rm fluid}=0.4$, 
indicating that particles get jammed in a position 
that is actually very close to their original position in the parent fluid.
This observation is of course consistent with the fact that the 
parent fluid and its corresponding jammed state display very similar 
geometric structure when $\phij$ is large, as shown above in 
Figs.~\ref{fig:Q_and_LFS}(b, d). 
Simultaneously, the finite-size effect on the tail in $P(\mathit{\Delta} r)$ 
almost disappears, which parallels the suppression of the
finite-size effect in $\phij$ values.  
Furthermore, $P(\mathit{\Delta} r)$ is no longer Gaussian, but 
exhibits an exponential tail, $P(\mathit{\Delta} r) 
\propto \exp[ - a \mathit{\Delta} r]$, where $a$ is a constant.
This functional form has been reported in supercooled liquids 
undergoing a transition between nearby inherent 
structures~\cite{schroder2000crossover,vogel2004particle,bailey2009exponential}.
This may suggest that during the compression of the denser fluid, 
the system may perform transitions among nearby locally stable 
configurations, while remaining firmly localised within a well-defined 
metabasin~\cite{charbonneau2015numerical,berthier2015growing,seguin2016experimental,jin2016exploring}.

\section{Discussion and conclusions}
\label{sec:conclusions}

Thanks to an efficient thermalisation algorithm which allowed us to
equilibrate polydisperse hard spheres up to unprecedented 
packing fractions~\cite{berthier2016equilibrium}, 
we have significantly extended the study of the
line of critical jamming transitions, or J-line. Our results
demonstrate that isostaticity and the associated critical behavior of
the pair correlation function remain unchanged along the entire J-line,
while some other structural properties at larger length scales evolve
qualitatively. Therefore, our results disentangle the structural
properties which originate from isostaticity at the contact scale from
other geometric properties at larger scale that seem unrelated or
insensitive to the jamming criticality.

We found that the fraction of rattlers in jammed packings increases
markedly with increasing $\phi_{\rm J}$, which is a counterintuitive
result. We confirm that a proper treatment of these rattlers is
essential to reveal the critical behavior at contact scale. Rattlers
also affect the organization of the packings at the neighbor-scale and tend to form relatively compact clusters, which
might induce subtle volume fraction fluctuations in the packing. In
Ref.~\cite{atkinson2016critical}, it has been argued that the presence of rattlers may interfere with hyperuniformity.
We found that deviations from hyperuniform behavior in jammed polydisperse hard spheres become more pronounced as the volume fraction increases.
However, hyperuniformity cannot be simply achieved by reducing the number of rattlers through additional compressions~\cite{wu2015search}.
Moreover, hyperuniform behavior in monodisperse packings disappears completely when rattlers are removed from the computation of the spectral density, or volume fraction fluctuation~\cite{donev2005unexpected}. Our results confirm this trend, which is in stark contrast with the jamming critical behavior seen in the pair correlation function.
This would be another supporting evidence that hyperuniformity and jamming criticality are unrelated.
These numerical results does not exclude the possibility that an ideal, rattler-free packing
obtained through a more complex optimization process provides
true hyperuniformity~\cite{atkinson2016critical}. Along this line of thought, it
would be interesting to study in more detail the relation between presence of rattlers 
and the violation of the hyperuniformity reported here.

We found that, above the onset volume fraction, the structure of jammed packings at the local scale closely  tracks the one of the parent equilibrium fluid. This contrasts with the behavior at the contact scale, which is qualitatively 
very different in jammed and fluid states.
In particular, as the volume fraction increases, both equilibrium fluid and jammed packings display a smooth but progressive local ordering, which we attributed to the emergence of distorted icosahedral structures. We attributed their irregular shape to the significant size dispersity of the system.
We speculate that only the most regular structures, identified through the asphericity parameter used in this work or more sophisticated metrics~\cite{Mickel_Kapfer_Schroder-Turk_Mecke_2013}, may actually provide locally stable arrangements in the dense equilibrium fluid.
These ideas might find useful applications in the context of glass structure studies~\cite{Royall_Williams_2015} and of investigations of the structure-dynamics relationship~\cite{hocky_correlation_2014}.

In a recent numerical study of polydisperse hard spheres~\cite{berthier2017breaking}, we demonstrated the growth of non-trivial static ``point-to-set'' correlations, which are at the core of thermodynamic pictures of the glass transition~\cite{BB04}.
It has been suggested that such ``amorphous order'' could be identified with hyperuniformity~\cite{hopkins2012nonequilibrium,coniglio2017universal}, although this link has been questioned~\cite{ikeda2015thermal}.
Our results cast further doubts on this argument: point-to-set correlations grow as the volume fraction increases, while hyperuniform behavior is suppressed.
The growth of amorphous order in the equilibrium fluid appears thus to compete with, rather than to enhance, hyperuniformity.

The significant suppression of finite size effects 
at higher $\phi_{\rm J}$ suggests
that the length scale causing this finite-size effect might not 
be related to the isostatic nature of the jamming transition, 
since all studied configurations are similarly isostatic.
This conclusion seems to contrast with other 
diverging length scales near jamming~\cite{silbert2005vibrations,goodrich2013stability,schoenholz2013stability,ellenbroek2009jammed,ikeda2013dynamic,ikeda2015thermal}.
We notice that the finite-size effect for inherent structure energies 
in Lennard-Jones supercooled liquids is also suppressed when considering 
lower temperatures for the parent fluid~\cite{doliwa2003finite},
which is reminiscent of our observations for 
denser $\phi_{\rm fluid}$ in hard spheres.
Finite size effects for jammed and inherent structures 
may have a common physical origin, namely the different topography of
the energy landscape probed in dilute or dense fluids~\cite{mari2009jamming,kim2003glass}. Dense fluids 
jam in a configuration that belongs
to well-defined metabasins, whereas dilute fluids may explore
a larger portion of the potential energy landscape before jamming,
larger systems exploring a larger region of phase space.

We have used a system with continuous polydispersity to enhance thermalisation.
However, we expect the conclusions drawn from the present model to hold more generally.
Systems with discrete polydispersity, \textit{i.e.} mixtures, are 
more easily amenable to theoretical investigations~\cite{parisi2010mean,baule2016edwards,ogarko2012equation,ogarko2013prediction}.
Recently, it has been proposed that thermal glassy systems with continuous 
polydispersity can be mapped into an effective multi-component system to characterize their thermodynamic behavior~\cite{ozawa2016does}.
Whether such an effective description is also applicable to the structural characterization of jammed states is an interesting open issue, which will likely require a careful treatment of the rattlers.~\cite{ogarko2012equation,ogarko2013prediction}.   

\begin{acknowledgments}
We thank D. Durian, P. Charbonneau, A. Ikeda, H. Ikeda, K. Kim, A. Liu, A. Ninarello, 
M. Pica Ciamarra, P. Sollich, and D. V{\aa}gberg for helpful discussions.
The numerical simulations were partially performed at Research 
Center of Computational Science (RCCS),
Okazaki, Japan.
We thank K. Kim for providing us with CPU time in RCCS.
The research leading to
these results has received funding from the European Research
Council under the European Unions Seventh Framework 
Programme (No. FP7/2007-2013)/ERC Grant Agreement No.306845. 
This work was supported by a grant from the Simons
Foundation (No. 454933, Ludovic Berthier).
\end{acknowledgments}

% This is necessary to redefine \doi with revtex, the one in the bst file is ineffective
\renewcommand{\doi}[1]{\href{http://dx.doi.org/#1}{doi:\discretionary{}{}{}#1}}
\bibliographystyle{SciPost_bibstyle}
\bibliography{paper}

\begin{thebibliography}{100}
\providecommand{\url}[1]{\texttt{#1}}
\providecommand{\urlprefix}{}
\expandafter\ifx\csname urlstyle\endcsname\relax
  \providecommand{\doi}[1]{doi:\discretionary{}{}{}#1}\else
  \providecommand{\doi}{doi:\discretionary{}{}{}\begingroup
  \urlstyle{rm}\Url}\fi
\providecommand{\eprint}[2][]{\url{#2}}

\bibitem{liu1998nonlinear}
A.~J. Liu and S.~R. Nagel,
\newblock \emph{Nonlinear dynamics: Jamming is not just cool any more},
\newblock Nature \textbf{396}, 21 (1998),
\newblock \doi{10.1038/23819}.

\bibitem{torquato2010jammed}
S.~Torquato and F.~H. Stillinger,
\newblock \emph{Jammed hard-particle packings: From kepler to bernal and
  beyond},
\newblock Rev. Mod. Phys. \textbf{82}, 2633 (2010),
\newblock \doi{10.1103/RevModPhys.82.2633}.

\bibitem{van2009jamming}
M.~Van~Hecke,
\newblock \emph{Jamming of soft particles: geometry, mechanics, scaling and
  isostaticity},
\newblock J. Phys.: Condens. Matter \textbf{22}, 033101 (2009),
\newblock \doi{10.1088/0953-8984/22/3/033101}.

\bibitem{liu2010jamming}
A.~J. Liu and S.~R. Nagel,
\newblock \emph{The jamming transition and the marginally jammed solid},
\newblock Annu. Rev. Condens. Matter Phys. \textbf{1}, 347 (2010),
\newblock \doi{10.1146/annurev-conmatphys-070909-104045}.

\bibitem{o2003jamming}
C.~S. O'Hern, L.~E. Silbert, A.~J. Liu and S.~R. Nagel,
\newblock \emph{Jamming at zero temperature and zero applied stress: The
  epitome of disorder},
\newblock Phys. Rev. E \textbf{68}, 011306 (2003),
\newblock \doi{10.1103/PhysRevE.68.011306}.

\bibitem{wyart2005effects}
M.~Wyart, L.~E. Silbert, S.~R. Nagel and T.~A. Witten,
\newblock \emph{Effects of compression on the vibrational modes of marginally
  jammed solids},
\newblock Phys. Rev. E \textbf{72}, 051306 (2005),
\newblock \doi{10.1103/PhysRevE.72.051306}.

\bibitem{olsson2007critical}
P.~Olsson and S.~Teitel,
\newblock \emph{Critical scaling of shear viscosity at the jamming transition},
\newblock Phys. Rev. Lett. \textbf{99}, 178001 (2007),
\newblock \doi{10.1103/PhysRevLett.99.178001}.

\bibitem{lerner2012unified}
E.~Lerner, G.~D{\"u}ring and M.~Wyart,
\newblock \emph{A unified framework for non-{Brownian} suspension flows and
  soft amorphous solids},
\newblock Proc. Natl. Acad. Sci. \textbf{109}, 4798 (2012),
\newblock \doi{10.1073/pnas.1120215109}.

\bibitem{donev2005pair}
A.~Donev, S.~Torquato and F.~H. Stillinger,
\newblock \emph{Pair correlation function characteristics of nearly jammed
  disordered and ordered hard-sphere packings},
\newblock Phys. Rev. E \textbf{71}, 011105 (2005),
\newblock \doi{10.1103/PhysRevE.71.011105}.

\bibitem{charbonneau2012universal}
P.~Charbonneau, E.~I. Corwin, G.~Parisi and F.~Zamponi,
\newblock \emph{Universal microstructure and mechanical stability of jammed
  packings},
\newblock Phys. Rev. Lett. \textbf{109}, 205501 (2012),
\newblock \doi{10.1103/PhysRevLett.109.205501}.

\bibitem{goodrich2016scaling}
C.~P. Goodrich, A.~J. Liu and J.~P. Sethna,
\newblock \emph{Scaling ansatz for the jamming transition},
\newblock Proc. Natl. Acad. Sci. \textbf{113}, 9745 (2016),
\newblock \doi{10.1073/pnas.1601858113}.

\bibitem{clusel2009granocentric}
M.~Clusel, E.~I. Corwin, A.~O. Siemens and J.~Bruji{\'c},
\newblock \emph{A 'granocentric' model for random packing of jammed emulsions},
\newblock Nature \textbf{460}, 611 (2009),
\newblock \doi{10.1038/nature08158}.

\bibitem{ogarko2014communication}
V.~Ogarko, N.~Rivas and S.~Luding,
\newblock \emph{Communication: Structure characterization of hard sphere
  packings in amorphous and crystalline states},
\newblock J. Chem. Phys. \textbf{140}, 211102 (2014),
\newblock \doi{10.1063/1.4880236}.

\bibitem{rieser2016divergence}
J.~M. Rieser, C.~P. Goodrich, A.~J. Liu and D.~J. Durian,
\newblock \emph{Divergence of voronoi cell anisotropy vector: A threshold-free
  characterization of local structure in amorphous materials},
\newblock Phys. Rev. Lett. \textbf{116}, 088001 (2016),
\newblock \doi{10.1103/PhysRevLett.116.088001}.

\bibitem{torquato2000random}
S.~Torquato, T.~M. Truskett and P.~G. Debenedetti,
\newblock \emph{Is random close packing of spheres well defined?},
\newblock Phys. Rev. Lett. \textbf{84}, 2064 (2000),
\newblock \doi{10.1103/PhysRevLett.84.2064}.

\bibitem{morse2014geometric}
P.~K. Morse and E.~I. Corwin,
\newblock \emph{Geometric signatures of jamming in the mechanical vacuum},
\newblock Phys. Rev. Lett. \textbf{112}, 115701 (2014),
\newblock \doi{10.1103/PhysRevLett.112.115701}.

\bibitem{Anikeenko_Medvedev_2007}
A.~V. Anikeenko and N.~N. Medvedev,
\newblock \emph{Polytetrahedral nature of the dense disordered packings of hard
  spheres},
\newblock Phys. Rev. Lett. \textbf{98}, 235504 (2007),
\newblock \doi{10.1103/PhysRevLett.98.235504}.

\bibitem{Kapfer_Mickel_Mecke_Schroder-Turk_2012}
S.~C. Kapfer, W.~Mickel, K.~Mecke and G.~E. Schr\"oder-Turk,
\newblock \emph{Jammed spheres: Minkowski tensors reveal onset of local
  crystallinity},
\newblock Phys. Rev. E \textbf{85}, 030301 (2012),
\newblock \doi{10.1103/PhysRevE.85.030301}.

\bibitem{Mickel_Kapfer_Schroder-Turk_Mecke_2013}
W.~Mickel, S.~C. Kapfer, G.~E. Schr\"oder-Turk and K.~Mecke,
\newblock \emph{Shortcomings of the bond orientational order parameters for the
  analysis of disordered particulate matter},
\newblock J. Chem. Phys. \textbf{138}, 044501 (2013),
\newblock \doi{10.1063/1.4774084}.

\bibitem{klatt2014characterization}
M.~A. Klatt and S.~Torquato,
\newblock \emph{Characterization of maximally random jammed sphere packings:
  Voronoi correlation functions},
\newblock Phys. Rev. E \textbf{90}, 052120 (2014),
\newblock \doi{10.1103/PhysRevE.90.052120}.

\bibitem{klumov2014structural}
B.~A. Klumov, Y.~Jin and H.~A. Makse,
\newblock \emph{Structural properties of dense hard sphere packings},
\newblock J. Phys. Chem. B \textbf{118}, 10761 (2014),
\newblock \doi{10.1021/jp504537n}.

\bibitem{donev2005unexpected}
A.~Donev, F.~H. Stillinger and S.~Torquato,
\newblock \emph{Unexpected density fluctuations in jammed disordered sphere
  packings},
\newblock Phys. Rev. Lett. \textbf{95}, 090604 (2005),
\newblock \doi{10.1103/PhysRevLett.116.088001}.

\bibitem{zachary2011hyperuniformity}
C.~E. Zachary, Y.~Jiao and S.~Torquato,
\newblock \emph{Hyperuniformity, quasi-long-range correlations, and void-space
  constraints in maximally random jammed particle packings. {I.} {Polydisperse}
  spheres},
\newblock Phys. Rev. E \textbf{83}, 051308 (2011),
\newblock \doi{10.1103/PhysRevE.83.051308}.

\bibitem{wu2015search}
Y.~Wu, P.~Olsson and S.~Teitel,
\newblock \emph{Search for hyperuniformity in mechanically stable packings of
  frictionless disks above jamming},
\newblock Phys. Rev. E \textbf{92}, 052206 (2015),
\newblock \doi{10.1103/PhysRevE.92.052206}.

\bibitem{ikeda2015thermal}
A.~Ikeda and L.~Berthier,
\newblock \emph{Thermal fluctuations, mechanical response, and hyperuniformity
  in jammed solids},
\newblock Phys. Rev. E \textbf{92}, 012309 (2015),
\newblock \doi{10.1103/PhysRevE.92.012309}.

\bibitem{ikedaparisi}
A.~Ikeda, L.~Berthier and G.~Parisi,
\newblock \emph{The large-scale structure of randomly jammed particles},
\newblock Phys. Rev. E \textbf{95}, 052125 (2017),
\newblock \doi{10.1103/PhysRevE.95.052125}.

\bibitem{speedy1998random}
R.~J. Speedy,
\newblock \emph{Random jammed packings of hard discs and spheres},
\newblock J. Phys.: Condens. Matter \textbf{10}, 4185 (1998),
\newblock \doi{10.1088/0953-8984/10/19/006}.

\bibitem{mari2009jamming}
R.~Mari, F.~Krzakala and J.~Kurchan,
\newblock \emph{Jamming versus glass transitions},
\newblock Phys. Rev. Lett. \textbf{103}, 025701 (2009),
\newblock \doi{10.1103/PhysRevLett.103.025701}.

\bibitem{biazzo2009theory}
I.~Biazzo, F.~Caltagirone, G.~Parisi and F.~Zamponi,
\newblock \emph{Theory of amorphous packings of binary mixtures of hard
  spheres},
\newblock Phys. Rev. Lett. \textbf{102}, 195701 (2009),
\newblock \doi{10.1103/PhysRevLett.102.195701}.

\bibitem{hermes2010jamming}
M.~Hermes and M.~Dijkstra,
\newblock \emph{Jamming of polydisperse hard spheres: The effect of kinetic
  arrest},
\newblock EPL (Europhysics Letters) \textbf{89}, 38005 (2010),
\newblock \doi{10.1209/0295-5075/89/38005}.

\bibitem{chaudhuri2010jamming}
P.~Chaudhuri, L.~Berthier and S.~Sastry,
\newblock \emph{Jamming transitions in amorphous packings of frictionless
  spheres occur over a continuous range of volume fractions},
\newblock Phys. Rev. Lett. \textbf{104}, 165701 (2010),
\newblock \doi{10.1103/PhysRevLett.104.165701}.

\bibitem{ciamarra2010recent}
M.~P. Ciamarra, M.~Nicodemi and A.~Coniglio,
\newblock \emph{Recent results on the jamming phase diagram},
\newblock Soft Matter \textbf{6}, 2871 (2010),
\newblock \doi{10.1039/B926810C}.

\bibitem{schreck2011tuning}
C.~F. Schreck, C.~S. O'Hern and L.~E. Silbert,
\newblock \emph{Tuning jammed frictionless disk packings from isostatic to
  hyperstatic},
\newblock Phys. Rev. E \textbf{84}, 011305 (2011),
\newblock \doi{10.1103/PhysRevE.84.011305}.

\bibitem{vaagberg2011glassiness}
D.~V{\aa}gberg, P.~Olsson and S.~Teitel,
\newblock \emph{Glassiness, rigidity, and jamming of frictionless soft core
  disks},
\newblock Phys. Rev. E \textbf{83}, 031307 (2011),
\newblock \doi{10.1103/PhysRevE.83.031307}.

\bibitem{otsuki2012critical}
M.~Otsuki and H.~Hayakawa,
\newblock \emph{Critical scaling of a jammed system after a quench of
  temperature},
\newblock Phys. Rev. E \textbf{86}, 031505 (2012),
\newblock \doi{10.1103/PhysRevE.86.031505}.

\bibitem{ozawa2012jamming}
M.~Ozawa, T.~Kuroiwa, A.~Ikeda and K.~Miyazaki,
\newblock \emph{Jamming transition and inherent structures of hard spheres and
  disks},
\newblock Phys. Rev. Lett. \textbf{109}, 205701 (2012),
\newblock \doi{10.1103/PhysRevLett.109.205701}.

\bibitem{baranau2014random}
V.~Baranau and U.~Tallarek,
\newblock \emph{Random-close packing limits for monodisperse and polydisperse
  hard spheres},
\newblock Soft Matter \textbf{10}, 3826 (2014),
\newblock \doi{0.1039/c3sm52959b}.

\bibitem{baranau2014jamming}
V.~Baranau and U.~Tallarek,
\newblock \emph{On the jamming phase diagram for frictionless hard-sphere
  packings},
\newblock Soft Matter \textbf{10}, 7838 (2014),
\newblock \doi{10.1039/C4SM01439A}.

\bibitem{bertrand2016protocol}
T.~Bertrand, R.~P. Behringer, B.~Chakraborty, C.~S. O'Hern and M.~D. Shattuck,
\newblock \emph{Protocol dependence of the jamming transition},
\newblock Phys. Rev. E \textbf{93}, 012901 (2016),
\newblock \doi{10.1103/PhysRevE.93.012901}.

\bibitem{kumar2016memory}
N.~Kumar and S.~Luding,
\newblock \emph{Memory of jamming--multiscale models for soft and granular
  matter},
\newblock Granular Matter \textbf{18}, 1 (2016),
\newblock \doi{10.1007/s10035-016-0624-2}.

\bibitem{luding2016so}
S.~Luding,
\newblock \emph{So much for the jamming point},
\newblock Nat. Phys. \textbf{12}, 531 (2016),
\newblock \doi{10.1038/nphys3680}.

\bibitem{witten09}
L.~Berthier and T.~A. Witten,
\newblock \emph{The glass transition of dense fluids of hard and compressible
  spheres},
\newblock Phys. Rev. E \textbf{80}, 021502 (2009),
\newblock \doi{10.1103/PhysRevE.80.021502}.

\bibitem{maiti2014}
M.~Maiti and S.~Sastry,
\newblock \emph{Free volume distribution of nearly jammed hard sphere
  packings},
\newblock J. Chem. Phys. \textbf{141}, 044510 (2014),
\newblock \doi{10.1063/1.4891358}.

\bibitem{berthier2016equilibrium}
L.~Berthier, D.~Coslovich, A.~Ninarello and M.~Ozawa,
\newblock \emph{Equilibrium sampling of hard spheres up to the jamming density
  and beyond},
\newblock Phys. Rev. Lett. \textbf{116}, 238002 (2016),
\newblock \doi{10.1103/PhysRevLett.116.238002}.

\bibitem{coslovich2007understanding}
D.~Coslovich and G.~Pastore,
\newblock \emph{Understanding fragility in supercooled {Lennard-Jones}
  mixtures. {I.} {Locally} preferred structures},
\newblock J. Chem. Phys. \textbf{127}, 124504 (2007),
\newblock \doi{10.1063/1.2773716}.

\bibitem{Royall_Williams_2015}
C.~P. Royall and S.~R. Williams,
\newblock \emph{The role of local structure in dynamical arrest},
\newblock Phys. Rep. \textbf{560}, 1 (2015),
\newblock \doi{10.1016/j.physrep.2014.11.004}.

\bibitem{atkinson2016critical}
S.~Atkinson, G.~Zhang, A.~B. Hopkins and S.~Torquato,
\newblock \emph{Critical slowing down and hyperuniformity on approach to
  jamming},
\newblock Phys. Rev. E \textbf{94}, 012902 (2016),
\newblock \doi{10.1103/PhysRevE.94.012902}.

\bibitem{allen1989computer}
M.~P. Allen and D.~J. Tildesley,
\newblock \emph{Computer simulation of liquids},
\newblock Oxford University Press, Oxford (1989).

\bibitem{santos2005contact}
A.~Santos, S.~B. Yuste and M.~L{\'o}pez~de Haro,
\newblock \emph{Contact values of the particle-particle and wall-particle
  correlation functions in a hard-sphere polydisperse fluid},
\newblock J. Chem. Phys. \textbf{123}(23), 234512 (2005),
\newblock \doi{doi:10.1063/1.2136883}.

\bibitem{ninarello2017}
A.~Ninarello, L.~Berthier and D.~Coslovich,
\newblock \emph{Models and algorithms for the next generation of glass
  transition studies},
\newblock Phys. Rev. X \textbf{7}, 021039 (2017),
\newblock \doi{10.1103/PhysRevX.7.021039}.

\bibitem{wilding2004phase}
N.~B. Wilding and P.~Sollich,
\newblock \emph{Phase equilibria and fractionation in a polydisperse fluid},
\newblock EPL (Europhysics Letters) \textbf{67}(2), 219 (2004),
\newblock \doi{10.1209/epl/i2004-10064-2}.

\bibitem{wilding2010phase}
N.~B. Wilding and P.~Sollich,
\newblock \emph{Phase behavior of polydisperse spheres: Simulation strategies
  and an application to the freezing transition},
\newblock J. Chem. Phys. \textbf{133}(22), 224102 (2010),
\newblock \doi{doi:10.1063/1.3510534}.

\bibitem{sindzingre1989calculation}
P.~Sindzingre, C.~Massobrio, G.~Ciccotti and D.~Frenkel,
\newblock \emph{Calculation of partial enthalpies of an argon-krypton mixture
  by {NPT} molecular dynamics},
\newblock Chem. Phys. \textbf{129}, 213 (1989),
\newblock \doi{10.1016/0301-0104(89)80007-2}.

\bibitem{gazzillo1989equation}
D.~Gazzillo and G.~Pastore,
\newblock \emph{Equation of state for symmetric non-additive hard-sphere
  fluids: An approximate analytic expression and new monte carlo results},
\newblock Chem. Phys. Lett. \textbf{159}, 388 (1989),
\newblock \doi{10.1016/0009-2614(89)87505-0}.

\bibitem{santen2001liquid}
L.~Santen and W.~Krauth,
\newblock \emph{Liquid, glass and crystal in two-dimensional hard disks},
\newblock arXiv:cond-mat/0107459  (2001),
\newblock \urlprefix\url{http://arxiv.org/abs/cond-mat/0107459}.

\bibitem{grigera2001fast}
T.~S. Grigera and G.~Parisi,
\newblock \emph{Fast {Monte} {Carlo} algorithm for supercooled soft spheres},
\newblock Phys. Rev. E \textbf{63}, 045102 (2001),
\newblock \doi{10.1103/PhysRevE.63.045102}.

\bibitem{brumer2004numerical}
Y.~Brumer and D.~R. Reichman,
\newblock \emph{Numerical investigation of the entropy crisis in model glass
  formers},
\newblock J. Phys. Chem. B \textbf{108}, 6832 (2004),
\newblock \doi{10.1021/jp037617y}.

\bibitem{fernandez2007phase}
L.~Fern{\'a}ndez, V.~Martin-Mayor and P.~Verrocchio,
\newblock \emph{Phase diagram of a polydisperse soft-spheres model for liquids
  and colloids},
\newblock Phys. Rev. Lett. \textbf{98}, 085702 (2007),
\newblock \doi{10.1103/PhysRevLett.98.085702}.

\bibitem{gutierrez2015static}
R.~Guti{\'e}rrez, S.~Karmakar, Y.~G. Pollack and I.~Procaccia,
\newblock \emph{The static lengthscale characterizing the glass transition at
  lower temperatures},
\newblock EPL (Europhysics Letters) \textbf{111}, 56009 (2015),
\newblock \doi{10.1209/0295-5075/111/56009}.

\bibitem{xu2005random}
N.~Xu, J.~Blawzdziewicz and C.~S. O'Hern,
\newblock \emph{Random close packing revisited: Ways to pack frictionless
  disks},
\newblock Phys. Rev. E \textbf{71}, 061306 (2005),
\newblock \doi{10.1103/PhysRevE.71.061306}.

\bibitem{desmond2009random}
K.~W. Desmond and E.~R. Weeks,
\newblock \emph{Random close packing of disks and spheres in confined
  geometries},
\newblock Phys. Rev. E \textbf{80}, 051305 (2009),
\newblock \doi{10.1103/PhysRevE.80.051305}.

\bibitem{nocedal1999numerical}
J.~Nocedal and S.~Wright,
\newblock \emph{Numerical Optimization},
\newblock Springer verlag, Berlin (1999).

\bibitem{dagois2012soft}
S.~Dagois-Bohy, B.~P. Tighe, J.~Simon, S.~Henkes and M.~van Hecke,
\newblock \emph{Soft-sphere packings at finite pressure but unstable to shear},
\newblock Phys. Rev. Lett. \textbf{109}, 095703 (2012),
\newblock \doi{10.1103/PhysRevLett.109.095703}.

\bibitem{berthier2015growing}
L.~Berthier, P.~Charbonneau, Y.~Jin, G.~Parisi, B.~Seoane and F.~Zamponi,
\newblock \emph{Growing timescales and lengthscales characterizing vibrations
  of amorphous solids},
\newblock Proc. Natl. Acad. Sci USA \textbf{113}, 8397 (2016),
\newblock \doi{10.1073/pnas.1607730113}.

\bibitem{desmond2014influence}
K.~W. Desmond and E.~R. Weeks,
\newblock \emph{Influence of particle size distribution on random close packing
  of spheres},
\newblock Phys. Rev. E \textbf{90}, 022204 (2014),
\newblock \doi{10.1103/PhysRevE.90.022204}.

\bibitem{sastry1998signatures}
S.~Sastry, P.~G. Debenedetti and F.~H. Stillinger,
\newblock \emph{Signatures of distinct dynamical regimes in the energy
  landscape of a glass-forming liquid},
\newblock Nature \textbf{393}, 554 (1998),
\newblock \doi{10.1038/31189}.

\bibitem{brumer2004mean}
Y.~Brumer and D.~R. Reichman,
\newblock \emph{Mean-field theory, mode-coupling theory, and the onset
  temperature in supercooled liquids},
\newblock Phys. Rev. E \textbf{69}, 041202 (2004),
\newblock \doi{10.1103/PhysRevE.69.041202}.

\bibitem{muller2015marginal}
M.~M{\"u}ller and M.~Wyart,
\newblock \emph{Marginal stability in structural, spin, and electron glasses},
\newblock Annu. Rev. Cond. Matt. Phys. \textbf{6}, 177 (2015),
\newblock \doi{10.1146/annurev-conmatphys-031214-014614}.

\bibitem{zhang2015structure}
C.~Zhang, C.~B. O'Donovan, E.~I. Corwin, F.~Cardinaux, T.~G. Mason, M.~E.
  M{\"o}bius and F.~Scheffold,
\newblock \emph{Structure of marginally jammed polydisperse packings of
  frictionless spheres},
\newblock Phys. Rev. E \textbf{91}, 032302 (2015),
\newblock \doi{10.1103/PhysRevE.91.032302}.

\bibitem{charbonneau2014fractal}
P.~Charbonneau, J.~Kurchan, G.~Parisi, P.~Urbani and F.~Zamponi,
\newblock \emph{Fractal free energy landscapes in structural glasses},
\newblock Nat. Comm. \textbf{5} (2014),
\newblock \doi{10.1038/ncomms4725}.

\bibitem{silbert2006structural}
L.~E. Silbert, A.~J. Liu and S.~R. Nagel,
\newblock \emph{Structural signatures of the unjamming transition at zero
  temperature},
\newblock Phys. Rev. E \textbf{73}, 041304 (2006),
\newblock \doi{10.1103/PhysRevE.73.041304}.

\bibitem{lerner2013low}
E.~Lerner, G.~D{\"u}ring and M.~Wyart,
\newblock \emph{Low-energy non-linear excitations in sphere packings},
\newblock Soft Matter \textbf{9}, 8252 (2013),
\newblock \doi{10.1039/C3SM50515D}.

\bibitem{charbonneau2015jamming}
P.~Charbonneau, E.~I. Corwin, G.~Parisi and F.~Zamponi,
\newblock \emph{Jamming criticality revealed by removing localized buckling
  excitations},
\newblock Phys. Rev. Lett. \textbf{114}, 125504 (2015),
\newblock \doi{10.1103/physrevlett.114.125504}.

\bibitem{steinhardt1983bond}
P.~J. Steinhardt, D.~R. Nelson and M.~Ronchetti,
\newblock \emph{Bond-orientational order in liquids and glasses},
\newblock Phys. Rev. B \textbf{28}, 784 (1983),
\newblock \doi{10.1103/PhysRevB.28.784}.

\bibitem{Moroni_2005}
D.~Moroni, P.~R. ten Wolde and P.~G. Bolhuis,
\newblock \emph{Interplay between structure and size in a critical crystal
  nucleus},
\newblock Phys. Rev. Lett. \textbf{94}, 235703 (2005),
\newblock \doi{10.1103/PhysRevLett.94.235703}.

\bibitem{Gellatly_Finney_1982}
B.~J. Gellatly and J.~L. Finney,
\newblock \emph{Characterisation of models of multicomponent amorphous metals:
  The radical alternative to the {Voronoi} polyhedron},
\newblock J. Non-Cryst. Solids \textbf{50}, 313 (1982),
\newblock \doi{10.1016/0022-3093(82)90093-X}.

\bibitem{voro++}
C.~H. Rycroft,
\newblock \emph{Voro++: A three-dimensional voronoi cell library in c++},
\newblock Chaos \textbf{19}, 041111 (2009),
\newblock \doi{10.1063/1.3215722}.

\bibitem{yamchi2015inherent}
M.~Z. Yamchi, S.~Ashwin and R.~K. Bowles,
\newblock \emph{Inherent structures, fragility, and jamming: Insights from
  quasi-one-dimensional hard disks},
\newblock Phys. Rev. E \textbf{91}, 022301 (2015),
\newblock \doi{10.1103/PhysRevE.91.022301}.

\bibitem{berthier2017breaking}
L.~Berthier, P.~Charbonneau, D.~Coslovich, A.~Ninarello, M.~Ozawa and S.~Yaida,
\newblock \emph{Breaking the glass ceiling: Configurational entropy
  measurements in extremely supercooled liquids},
\newblock arXiv:1704.08257  (2017),
\newblock \urlprefix\url{http://arxiv.org/1704.08257}.

\bibitem{lechner_accurate_2008}
W.~Lechner and C.~Dellago,
\newblock \emph{Accurate determination of crystal structures based on averaged
  local bond order parameters},
\newblock J. Chem. Phys. \textbf{129}, 114707 (2008),
\newblock \doi{10.1063/1.2977970}.

\bibitem{Finney_1976}
J.~L. Finney,
\newblock \emph{Fine structure in randomly packed, dense clusters of hard
  spheres},
\newblock Mat. Sci. Eng. \textbf{23}, 199 (1976),
\newblock \doi{10.1016/0025-5416(76)90194}.

\bibitem{Tanemura_Hiwatari_Matsuda_Ogawa_Ogita_Ueda_1977}
M.~Tanemura, Y.~Hiwatari, H.~Matsuda, T.~Ogawa, N.~Ogita and A.~Ueda,
\newblock \emph{Geometrical analysis of crystallization pf the soft-core
  model},
\newblock Prog. Theor. Phys. \textbf{58}, 1079 (1977),
\newblock \doi{10.1143/PTP.58.1079}.

\bibitem{Ma_2015}
E.~Ma,
\newblock \emph{Tuning order in disorder},
\newblock Nat. Mater. \textbf{14}, 547 (2015),
\newblock \doi{10.1038/nmat4300}.

\bibitem{Note1}
We disregard null values of $n_q$ for $q>q^\prime $, where $q^\prime $ is the
  largest number of vertices such that $n_q^\prime >0$.

\bibitem{malins_identification_2013}
A.~Malins, J.~Eggers, C.~P. Royall, S.~R. Williams and H.~Tanaka,
\newblock \emph{Identification of long-lived clusters and their link to slow
  dynamics in a model glass former},
\newblock J. Chem. Phys. \textbf{138}, 12A535 (2013),
\newblock \doi{10.1063/1.4790515}.

\bibitem{wahnstrom}
G.~Wahnstrom,
\newblock \emph{Molecular-dynamics study of a supercooled two-component
  {Lennard-Jones} system},
\newblock Phys. Rev. A \textbf{44}, 3752 (1991),
\newblock \doi{10.1103/PhysRevA.44.3752}.

\bibitem{coslovich2011locally}
D.~Coslovich,
\newblock \emph{Locally preferred structures and many-body static correlations
  in viscous liquids},
\newblock Phys. Rev. E \textbf{83}, 051505 (2011),
\newblock \doi{10.1103/PhysRevE.83.051505}.

\bibitem{leocmach2012roles}
M.~Leocmach and H.~Tanaka,
\newblock \emph{Roles of icosahedral and crystal-like order in the hard spheres
  glass transition},
\newblock Nat. Comm. \textbf{3}, 974 (2012),
\newblock \doi{10.1038/ncomms1974}.

\bibitem{Dunleavy_Wiesner_Yamamoto_Royall_2015}
A.~J. Dunleavy, K.~Wiesner, R.~Yamamoto and C.~P. Royall,
\newblock \emph{Mutual information reveals multiple structural relaxation
  mechanisms in a model glass former},
\newblock Nat. Comm. \textbf{6}, 6089 (2015),
\newblock \doi{10.1038/ncomms7089}.

\bibitem{Schroder-Turk_Mickel_Schroter_Delaney_Saadatfar_Senden_Mecke_Aste_2010}
G.~E. Schr{\"o}der-Turk, W.~Mickel, M.~Schr{\"o}ter, G.~W. Delaney,
  M.~Saadatfar, T.~J. Senden, K.~Mecke and T.~Aste,
\newblock \emph{Disordered spherical bead packs are anisotropic},
\newblock EPL (Europhysics Letters) \textbf{90}, 34001 (2010),
\newblock \doi{10.1209/0295-5075/90/34001}.

\bibitem{hocky_correlation_2014}
G.~M. Hocky, D.~Coslovich, A.~Ikeda and D.~R. Reichman,
\newblock \emph{Correlation of local order with particle mobility in
  supercooled liquids is highly system dependent},
\newblock Phys. Rev. Lett. \textbf{113}, 157801 (2014),
\newblock \doi{10.1103/PhysRevLett.113.157801}.

\bibitem{atkinson2013detailed}
S.~Atkinson, F.~H. Stillinger and S.~Torquato,
\newblock \emph{Detailed characterization of rattlers in exactly isostatic,
  strictly jammed sphere packings},
\newblock Phys. Rev. E \textbf{88}, 062208 (2013),
\newblock \doi{10.1103/PhysRevE.88.062208}.

\bibitem{zachary2011hyperuniformPRL}
C.~E. Zachary, Y.~Jiao and S.~Torquato,
\newblock \emph{Hyperuniform long-range correlations are a signature of
  disordered jammed hard-particle packings},
\newblock Phys. Rev. Lett. \textbf{106}, 178001 (2011),
\newblock \doi{10.1103/PhysRevLett.106.178001}.

\bibitem{torquato2003local}
S.~Torquato and F.~H. Stillinger,
\newblock \emph{Local density fluctuations, hyperuniformity, and order
  metrics},
\newblock Phys. Rev. E \textbf{68}, 041113 (2003),
\newblock \doi{10.1103/PhysRevE.68.041113}.

\bibitem{xu2010effects}
N.~Xu and E.~S. Ching,
\newblock \emph{Effects of particle-size ratio on jamming of binary mixtures at
  zero temperature},
\newblock Soft Matter \textbf{6}, 2944 (2010),
\newblock \doi{10.1039/b926696h}.

\bibitem{kurita2010experimental}
R.~Kurita and E.~R. Weeks,
\newblock \emph{Experimental study of random-close-packed colloidal particles},
\newblock Phys. Rev. E \textbf{82}, 011403 (2010),
\newblock \doi{10.1103/PhysRevE.82.011403}.

\bibitem{berthier2011suppressed}
L.~Berthier, P.~Chaudhuri, C.~Coulais, O.~Dauchot and P.~Sollich,
\newblock \emph{Suppressed compressibility at large scale in jammed packings of
  size-disperse spheres},
\newblock Phys. Rev. Lett. \textbf{106}, 120601 (2011),
\newblock \doi{10.1103/PhysRevLett.106.120601}.

\bibitem{klatt2016characterization}
M.~A. Klatt and S.~Torquato,
\newblock \emph{Characterization of maximally random jammed sphere packings.
  {II.} {Correlation} functions and density fluctuations},
\newblock Phys. Rev. E \textbf{94}, 022152 (2016),
\newblock \doi{10.1103/PhysRevE.94.022152}.

\bibitem{binder2011glassy}
K.~Binder and W.~Kob,
\newblock \emph{Glassy materials and disordered solids: An introduction to
  their statistical mechanics},
\newblock World Scientific (2011).

\bibitem{hopkins2012nonequilibrium}
A.~B. Hopkins, F.~H. Stillinger and S.~Torquato,
\newblock \emph{Nonequilibrium static diverging length scales on approaching a
  prototypical model glassy state},
\newblock Phys. Rev. E \textbf{86}, 021505 (2012),
\newblock \doi{10.1103/PhysRevE.86.021505}.

\bibitem{ikeda2013dynamic}
A.~Ikeda, L.~Berthier and G.~Biroli,
\newblock \emph{Dynamic criticality at the jamming transition},
\newblock J. Chem. Phys. \textbf{138}, 12A507 (2013),
\newblock \doi{10.1063/1.4769251}.

\bibitem{parisi2010mean}
G.~Parisi and F.~Zamponi,
\newblock \emph{Mean-field theory of hard sphere glasses and jamming},
\newblock Rev. Mod. Phys. \textbf{82}, 789 (2010),
\newblock \doi{10.1103/RevModPhys.82.789}.

\bibitem{coniglio2017universal}
A.~Coniglio, M.~P. Ciamarra and T.~Aste,
\newblock \emph{Universal behaviour of the glass and the jamming transitions in
  finite dimensions},
\newblock arXiv:1704.01231  (2017),
\newblock \urlprefix\url{http://arxiv.org/abs/1704.01231}.

\bibitem{nishimori2010elements}
H.~Nishimori and G.~Ortiz,
\newblock \emph{Elements of phase transitions and critical phenomena},
\newblock Oxford University Press, Oxford (2010).

\bibitem{silbert2005vibrations}
L.~E. Silbert, A.~J. Liu and S.~R. Nagel,
\newblock \emph{Vibrations and diverging length scales near the unjamming
  transition},
\newblock Phys. Rev. Lett. \textbf{95}, 098301 (2005),
\newblock \doi{10.1103/PhysRevLett.95.098301}.

\bibitem{drocco2005multiscaling}
J.~Drocco, M.~Hastings, C.~O. Reichhardt and C.~Reichhardt,
\newblock \emph{Multiscaling at point {J}: Jamming is a critical phenomenon},
\newblock Phys. Rev. Lett. \textbf{95}, 088001 (2005),
\newblock \doi{10.1103/PhysRevLett.95.088001}.

\bibitem{reichhardt2014aspects}
C.~Reichhardt and C.~O. Reichhardt,
\newblock \emph{Aspects of jamming in two-dimensional athermal frictionless
  systems},
\newblock Soft Matter \textbf{10}, 2932 (2014),
\newblock \doi{10.1039/c3sm53154f}.

\bibitem{goodrich2013stability}
C.~P. Goodrich, W.~G. Ellenbroek and A.~J. Liu,
\newblock \emph{Stability of jammed packings {I}: the rigidity length scale},
\newblock Soft Matter \textbf{9}, 10993 (2013),
\newblock \doi{10.1039/C3SM51095F}.

\bibitem{schoenholz2013stability}
S.~S. Schoenholz, C.~P. Goodrich, O.~Kogan, A.~J. Liu and S.~R. Nagel,
\newblock \emph{Stability of jammed packings {II}: the transverse length
  scale},
\newblock Soft Matter \textbf{9}, 11000 (2013),
\newblock \doi{10.1039/C3SM51096D}.

\bibitem{ellenbroek2009jammed}
W.~G. Ellenbroek, M.~van Hecke and W.~van Saarloos,
\newblock \emph{Jammed frictionless disks: Connecting local and global
  response},
\newblock Phys. Rev. E \textbf{80}, 061307 (2009),
\newblock \doi{10.1103/PhysRevE.80.061307}.

\bibitem{ikeda2015one}
H.~Ikeda and A.~Ikeda,
\newblock \emph{One-dimensional {KAC} model of dense amorphous hard spheres},
\newblock EPL (Europhysics Letters) \textbf{111}, 40007 (2015),
\newblock \doi{10.1209/0295-5075/111/40007}.

\bibitem{vaagberg2011finite}
D.~V{\aa}gberg, D.~Valdez-Balderas, M.~Moore, P.~Olsson and S.~Teitel,
\newblock \emph{Finite-size scaling at the jamming transition: Corrections to
  scaling and the correlation-length critical exponent},
\newblock Phys. Rev. E \textbf{83}, 030303 (2011),
\newblock \doi{10.1103/PhysRevE.83.030303}.

\bibitem{gao2006frequency}
G.-J. Gao, J.~B{\l}awzdziewicz and C.~S. O'Hern,
\newblock \emph{Frequency distribution of mechanically stable disk packings},
\newblock Phys. Rev. E \textbf{74}, 061304 (2006),
\newblock \doi{10.1103/PhysRevE.74.061304}.

\bibitem{ciamarra2010disordered}
M.~P. Ciamarra, A.~Coniglio and A.~de~Candia,
\newblock \emph{Disordered jammed packings of frictionless spheres},
\newblock Soft Matter \textbf{6}, 2975 (2010),
\newblock \doi{10.1039/C001904F}.

\bibitem{schroder2000crossover}
T.~B. Schr{\o}der, S.~Sastry, J.~C. Dyre and S.~C. Glotzer,
\newblock \emph{Crossover to potential energy landscape dominated dynamics in a
  model glass-forming liquid},
\newblock J. Chem. Phys. \textbf{112}, 9834 (2000),
\newblock \doi{10.1063/1.481621}.

\bibitem{vogel2004particle}
M.~Vogel, B.~Doliwa, A.~Heuer and S.~C. Glotzer,
\newblock \emph{Particle rearrangements during transitions between local minima
  of the potential energy landscape of a binary {Lennard-Jones} liquid},
\newblock J. Chem. Phys. \textbf{120}, 4404 (2004),
\newblock \doi{10.1063/1.1644538}.

\bibitem{bailey2009exponential}
N.~P. Bailey, T.~B. Schr{\o}der and J.~C. Dyre,
\newblock \emph{Exponential distributions of collective flow-event properties
  in viscous liquid dynamics},
\newblock Phys. Rev. Lett. \textbf{102}, 055701 (2009),
\newblock \doi{10.1103/PhysRevLett.102.055701}.

\bibitem{charbonneau2015numerical}
P.~Charbonneau, Y.~Jin, G.~Parisi, C.~Rainone, B.~Seoane and F.~Zamponi,
\newblock \emph{Numerical detection of the {Gardner} transition in a mean-field
  glass former},
\newblock Phys. Rev. E \textbf{92}, 012316 (2015),
\newblock \doi{10.1103/PhysRevE.92.012316}.

\bibitem{seguin2016experimental}
A.~Seguin and O.~Dauchot,
\newblock \emph{Experimental evidences of the {Gardner} phase in a granular
  glass},
\newblock Phys. Rev. Lett. \textbf{117}, 228001 (2016),
\newblock \doi{10.1103/PhysRevLett.117.228001}.

\bibitem{jin2016exploring}
Y.~Jin and H.~Yoshino,
\newblock \emph{Exploring the complex free energy landscape of the simplest
  glass by rheology},
\newblock Nat. Comm. \textbf{8}, 14935 (2016),
\newblock \doi{10.1038/ncomms14935}.

\bibitem{BB04}
J.-P. Bouchaud and G.~Biroli,
\newblock \emph{On the {Adam}-{Gibbs}-{Kirkpatrick}-{Thirumalai}-{Wolynes}
  scenario for the viscosity increase in glasses},
\newblock J. Chem. Phys. \textbf{121}, 7347 (2004),
\newblock \doi{doi:10.1063/1.1796231}.

\bibitem{doliwa2003finite}
B.~Doliwa and A.~Heuer,
\newblock \emph{Finite-size effects in a supercooled liquid},
\newblock J. Phys.: Condens. Matter \textbf{15}, S849 (2003),
\newblock \doi{10.1088/0953-8984/15/11/309}.

\bibitem{kim2003glass}
K.~Kim and T.~Munakata,
\newblock \emph{Glass transition of hard sphere systems: Molecular dynamics and
  density functional theory},
\newblock Phys. Rev. E \textbf{68}(2), 021502 (2003),
\newblock \doi{10.1103/PhysRevE.68.021502}.

\bibitem{baule2016edwards}
A.~Baule, F.~Morone, H.~J. Herrmann and H.~A. Makse,
\newblock \emph{Edwards statistical mechanics for jammed granular matter},
\newblock arXiv:1602.04369  (2016),
\newblock \urlprefix\url{http://arxiv.org/abs/1602.04369}.

\bibitem{ogarko2012equation}
V.~Ogarko and S.~Luding,
\newblock \emph{Equation of state and jamming density for equivalent bi-and
  polydisperse, smooth, hard sphere systems},
\newblock J. Chem. Phys. \textbf{136}, 124508 (2012),
\newblock \doi{10.1063/1.3694030}.

\bibitem{ogarko2013prediction}
V.~Ogarko and S.~Luding,
\newblock \emph{Prediction of polydisperse hard-sphere mixture behavior using
  tridisperse systems},
\newblock Soft Matter \textbf{9}, 9530 (2013),
\newblock \doi{10.1039/C3SM50964H}.

\bibitem{ozawa2016does}
M.~Ozawa and L.~Berthier,
\newblock \emph{Does the configurational entropy of polydisperse particles
  exist?},
\newblock J. Chem. Phys. \textbf{146}, 014502 (2017),
\newblock \doi{10.1063/1.4972525}.

\end{thebibliography}

\end{document}